\documentclass[prd,showpacs,superscriptaddress,preprintnumbers,amsmath,nofootinbib]{revtex4}
\usepackage[dvips,final]{graphicx}
\usepackage{amssymb,amsmath,epsfig,bm,pifont,amsfonts}
\usepackage[utf8]{inputenc}
\newcommand{\be}{\begin{equation}}\newcommand{\ee}{\end{equation}}%
\newcommand{\bd}{\begin{displaymath}}\newcommand{\ed}{\end{displaymath}}
\newcommand{\bit}{\begin{itemize}}                                        
 \newcommand{\eit}{\end{itemize}}                                         
\newcommand{\ben}{\begin{enumerate}}                                      
 \newcommand{\een}{\end{enumerate}}                                       
\newcommand{\baa}{\begin{array}{lll}}                                     
 \newcommand{\eaa}{\end{array}}                                           
\newcommand{\ba}{\begin{eqnarray}}                                        
 \newcommand{\ea}{\end{eqnarray}}                                         
\newcommand{\gev}[1]{\relax\ifmmode{\text{GeV}^{#1}}                      
                     \else{GeV$^{#1}${ }}\fi}                             
\def\MSbar{\relax\ifmmode\overline                                        
            {\rm MS}\else{$\overline{\rm MS}${ }}\fi}                     
\def\as{\relax\ifmmode \alpha_s\else{$ \alpha_s${ }}\fi}                  
\def\abar{\relax\ifmmode{\bar{a}}\else{$\bar{a}${ }}\fi}                  
  \def\ie{\hbox{\it i.e.}{ }}                 
    
\usepackage{color}
\definecolor{mBlue}{rgb}{0,0,1}
 
\definecolor{mRed}{rgb}{1,0,0}
 
\definecolor{mGreen}{rgb}{0.0,1.0,1.0}
 
    \definecolor{DarkGreen}{rgb}{0.04,0.5,0.1}

\begin{document}
\thispagestyle{empty}
 \date{\today}

\title{Evaluation of the Gottfried sum with use of the truncated moments method}



\author{A. Kotlorz}
\email{a.kotlorz@po.edu.pl}
\affiliation{Opole University of Technology, 45-758 Opole, Proszkowska 76, Poland}

\author{D. Kotlorz}
\email{dorota@theor.jinr.ru}
\affiliation{Opole University of Technology, 45-758 Opole, Proszkowska 76, Poland}
\affiliation{Bogoliubov Laboratory of Theoretical Physics, JINR, 141980 Dubna, Russia}

\author{O. V. Teryaev}
\email{teryaev@theor.jinr.ru}
\affiliation{Bogoliubov Laboratory of Theoretical Physics, JINR, 141980 Dubna, Russia}


\begin{abstract}
We reanalyze the experimental NMC data on the nonsinglet structure function $F_2^p-F_2^n$
and E866 data on the nucleon sea asymmetry $\bar{d}/\bar{u}$ using the truncated
moments approach elaborated in our previous papers. With help of the special truncated sum
one can overcome the problem of the unavoidable experimental restrictions on the Bjorken $x$
and effectively study the fundamental sum rules for the parton distributions and structure
functions.
Using only the data from the measured region of $x$, we obtain the
Gottfried sum $\int_0^1 F_2^{ns}/x\, dx$ and the integrated nucleon sea asymmetry
$\int_0^1 (\bar{d}-\bar{u})\, dx$.
We compare our results with the reported experimental values and with the
predictions obtained for different global parametrizations for the parton
distributions.
We also discuss the discrepancy between the NMC and E866 results
on $\int_0^1 (\bar{d}-\bar{u})\, dx$.
We demonstrate that this discrepancy can be resolved by taking into account
the higher-twist effects.
\end{abstract}
\pdfinfo{%
  /Title    ()
  /Author   ()
  /Creator  ()
  /Producer ()
  /Subject  ()
  /Keywords ()
}

\maketitle
\section{Introduction}
\label{sec:intro}
The deep inelastic scattering (DIS) of leptons on hadrons and hadron-hadron
collisions are a gold mine to study the hadron structure and fundamental
particle interactions at high energies. Especially, so-called DIS sum rules
can provide important information on partonic structure of the nucleon and
a good test for the quantum chromodynamics (QCD).
Nowadays, there are known a number of polarized and unpolarized sum rules
for structure functions. Some of them are rigorous theoretical predictions and
other are based on model assumptions which can be verified experimentally.
An example of the latter is the Gottfried sum rule (GSR) \cite{Gottfried:1967kk}.
Thus, the GSR violation in a series of experiments \cite{Amaudruz:1991at, Arneodo:1994sh,
Baldit:1994jk, Hawker:1998ty, Peng:1998pa, Towell:2001nh, Ackerstaff:1998sr}
revealed that, unlikely to the assumed simple partonic model of the
nucleon with the symmetric light sea, the light sea of the proton was flavor
asymmetric, i.e., $\bar{u}(x)\neq\bar{d}(x)$. This unexpected result has
prompted a large interest for many further studies, for review, see, e.g.,
\cite{Kumano:1997cy, Garvey:2001yq}, related to theoretical explanations of
the flavor asymmetry of the nucleon sea.

In our paper, we present a phenomenological analysis of the experimental
NMC data on the nonsinglet structure function $F_2^p-F_2^n$ \cite{Arneodo:1994sh}
and E866 data on the nucleon sea asymmetry $\bar{d}/\bar{u}$ \cite{Towell:2001nh},
utilizing a very effective method for determination of the DIS sum rules in
a restricted region of Bjorken $x$ -- the so-called truncated Mellin moments
(TMM) approach \cite{Kotlorz:2017wpu}.\\

In the next section, we give a brief recapitulation of the Gottfried sum rule
violation problem and discuss some effects modifying the GSR like the perturbative
QCD corrections, higher-twist terms, small-$x$ behavior and nuclear shadowing.
The method of the evaluation of the DIS rules from the experimental
data with help of the truncated Mellin moments approach is shortly
summarized in Section III.
In Section IV, we present our numerical results on the GSR value and compare
them to those provided by the NMC and E866, and also to other determinations based
on the global parton distribution functions (PDFs) fits.
Furthermore, we discuss the higher-twist effects as a possible explanation of
the discrepancy between the NMC and E866 results
on the integrated nucleon sea asymmetry $\int_0^1 (\bar{d}-\bar{u})\, dx$. 
Finally, we discuss shortly our prediction for the iso-vector quark
momentum fraction $\langle x \rangle_{u-d}$.\\
In Section~V, we give conclusions for this study.

\section{Violation of the Gottfried sum rule}
\label{sec:sec2}
The Gottfried sum rule \cite{Gottfried:1967kk} states that the integral over
Bjorken variable $0 < x < 1$ of a difference of electron-proton and
electron-neutron structure functions is a constant ($=1/3$) under flavor symmetry
in the nucleon sea ($\bar{u}(x) = \bar{d}(x)$), which is independent of the
transferred four-momentum $q$ \cite{Gottfried:1967kk}:
\begin{equation} \label{eq2.6}
S_G(Q^2) = \int^1_{0} \left [F_2^p(x,Q^2) - F_2^n(x,Q^2)\right ] {dx\over x}
= \frac{1}{3}\, .
\end{equation}
Here, $x = Q^2/(2Pq)$, where $Q^2=-q^2$,
$P^2=m^2$, and $m$ is the nucleon mass.
This form of the GSR originates from a simple partonic model of the nucleon
structure functions in which the isospin symmetry of the nucleon (the u-quark
distribution in the proton is equal to the d-quark distribution in the neutron),
\begin{equation}\label{eq2.1}
u^p_{v}(x) = d^n_{v}(x) \equiv u_{v}(x)\, ,
\end{equation}
and, similarly, $d^p = u^n$ , $\bar{u}^p = \bar{d}^n$, $\bar{d}^p = \bar{u}^n$, etc.,
and the flavor symmetry of the light sea in the nucleon,
\begin{equation}\label{eq2.2}
\bar{u}(x) = \bar{d}(x)\, ,
\end{equation}
are assumed.
Then, the difference between the proton and neutron structure functions
incorporating implicit perturbative QCD $Q^2$ corrections to the parton
model is given by
\begin{equation}\label{eq2.3}
F_2^p(x, Q^2) - F_2^n(x, Q^2)  = \frac{1}{3}\,x\left [ u_v(x,Q^2)
- d_v(x,Q^2)\right ] + \frac{2}{3}\,x\left [ \bar{u}(x,Q^2) -
\bar{d}(x,Q^2)\right ],
\end{equation}
where the valence-quark distribution $q_v$, ($q=u,d$), is defined by
$q_v\equiv q-\bar{q}$, with $\bar{q}$ being the sea-quark distribution.
Taking into account the charge conservation law for the nucleon,
\begin{equation}\label{eq2.4}
\int^1_{0} u_{v}(x,Q^2)\, dx = 2\, ,\quad\quad \int^1_{0} d_{v}(x,Q^2)\, dx = 1\, ,
\end{equation}
we obtain
\begin{equation}\label{eq2.5}
\int^1_{0} \left [F_2^p(x,Q^2) - F_2^n(x,Q^2)\right ]{dx\over x} =
\frac{1}{3} + \frac{2}{3}\int^1_{0} \left [ \bar{u}(x,Q^2) - \bar{d}(x,Q^2)\right ] dx.
\end{equation}
If the light sea is flavor symmetric, Eq.~(\ref{eq2.2}), the second term
in Eq.~(\ref{eq2.5}) vanishes giving the Gottfried sum rule (\ref{eq2.6}).

Though the isospin symmetry, Eq.~(\ref{eq2.1}), is not exact, and can also
contribute to the GSR violation, usually the experimental results on the GSR
breaking are interpreted as an evidence of the light flavor asymmetry
of the nucleon sea,
\begin{equation}\label{eq2.7}
\bar{u}(x) \neq \bar{d}(x)\, .
\end{equation}
The first clear indication of the GSR violation in DIS experiment was
provided by the New Muon Collaboration (NMC) \cite{Amaudruz:1991at} and from
the reanalyzed NMC data \cite{Arneodo:1994sh}.
The obtained NMC measurement of $S_G$,
\begin{equation}\label{eq2.8}
{\rm NMC~~1994:}\quad\quad S_G(Q^2=4\,{\rm GeV}^2) = 0.235\pm 0.026
\end{equation}
implies the integrated antiquark flavor asymmetry, Eq.~(\ref{eq2.7}),
\begin{equation}\label{eq2.9}
\int^1_{0} \left [ \bar{d}(x,Q^2) - \bar{u}(x,Q^2)\right ] dx = 0.148\pm 0.039
\end{equation}
which means that in the proton, $d$-sea is larger than $u$-sea.

Later, the Gottfried sum rule was tested at the Fermilab in E866 Drell-Yan (DY)
experiments which measured $\bar{d}/\bar{u}$ as a function of $x$ over the
kinematic range of $0.015<x<0.35$ at $Q^2=54\, {\rm GeV}^2$ \cite{Towell:2001nh}.
Again, the data suggested a significant deficit in the sum rule consistent with the
DIS results and also with semi-inclusive DIS (SIDIS) measurements of the
HERMES collaboration for $0.020<x<0.30$ and $1<Q^2<20\, {\rm GeV}^2$
\cite{Ackerstaff:1998sr}:
\begin{subequations}
\label{eq2.10}
\begin{eqnarray}
\label{eq2.10a}
{\rm HERMES~~1998:}\quad\quad
&&\int^1_{0} \left [ \bar{d}(x,Q^2)-\bar{u}(x,Q^2)\right ] dx = 0.16\pm 0.03 \\
\label{eq2.10b}
{\rm E866~~2001:}\quad\quad
&&\int^1_{0} \left [ \bar{d}(x,Q^2)-\bar{u}(x,Q^2)\right ] dx = 0.118\pm 0.012
\end{eqnarray}
\end{subequations}

The surprisingly large difference between light sea in the nucleon,
Eq.~(\ref{eq2.7}), observed in different experiments like DIS, DY and SIDIS,
has triggered many theoretical efforts to understand and accurately describe
the experimental results (for review, see, e.g., \cite{Kumano:1997cy, Garvey:2001yq}).
While the perturbative QCD fails in description of the sea asymmetry, the
nonperturbative mechanisms as Pauli-blocking, meson cloud, chiral-quark, intrinsic sea,
soliton seem to be more promising in explanation of the GSR breaking.
Recently, the statistical parton distributions approach was developed to study the
flavor structure of the light quark sea \cite{Soffer:2019gbb}. 
The authors obtained a remarkable agreement of the statistical model prediction
for the ratio $\bar{d}/\bar{u}$ with the E866 data \cite{Garvey:2001yq, Peng:2014uea}
up to $x=0.2$.
Unfortunately, none of the studies mentioned above predicts correctly the $\bar{d}/\bar{u}$
behavior in the whole $x$ region, i.e. none of them predicts a sign-change for
$\bar{d}(x)-\bar{u}(x)$ at $x\approx 0.3$ as suggested by the E866 data.\\
Below, we briefly discuss possible effects modifying the GSR like the perturbative QCD corrections,
higher twist-terms, small-$x$ behavior and nuclear shadowing effects.

\subsection{pQCD corrections to the GSR}
\label{sec:sec2.1}
Here, we show that the perturbative QCD corrections to the GSR are too small to explain the
light sea asymmetry \cite{Hinchliffe:1996hc}.
The corrections of order $\alpha_s^2$ to the GSR were obtained in \cite{Kataev:2003en}
basing on numerical calculation of the order $\alpha_s^2$ contribution to
the coefficient function.

From the renormalization group equation analysis for $S_G(Q^2)$  Kataev and
Parente \cite{Kataev:2003en} obtained for the number of active flavors
$n_f$=4 the following QCD corrections to the GSR:
\begin{equation}\label{eq2.12}
S_G(Q^2) = \frac{1}{3}\left [1+0.0384\left (\frac{\alpha_s}{\pi}\right ) -
0.822\left (\frac{\alpha_s}{\pi}\right )^2\right ].
\end{equation}
Using the above formula we find
\begin{subequations}
\label{eq2.13}
\begin{eqnarray}
\label{eq2.13a}
S_G(Q^2=4\,{\rm GeV}^2)  &=& \frac{1}{3} - 0.0015 = 0.3318\\
\label{eq2.13b}
S_G(Q^2=54\,{\rm GeV}^2) &=& \frac{1}{3} - 0.0003 = 0.3330\, .
\end{eqnarray}
\end{subequations}
This means that the magnitude of order $\alpha_s^2$ perturbative QCD
effects turn out to be about $-0.4\%$ at $Q^2=4\,{\rm GeV}^2$
($\alpha_s\approx 0.31$), and $-0.08\%$ at $Q^2=54\,{\rm GeV}^2$
($\alpha_s\approx 0.20$) of the original constant value of the GSR, $S_G=1/3$.\\
So, it is clearly seen that the perturbative QCD corrections to the Gottfried
sum rule are very small and cannot explain the experimental results of NMC
\cite{Arneodo:1994sh} and E866 \cite{Towell:2001nh} collaborations where the GSR
is broken on the level of $-29.5\%$ and $-23.6\%$, respectively.

\subsection{Higher-twist effects}
\label{sec:sec2.2}
\begin{figure}[ht]
\centering
\includegraphics[width=0.5\textwidth]{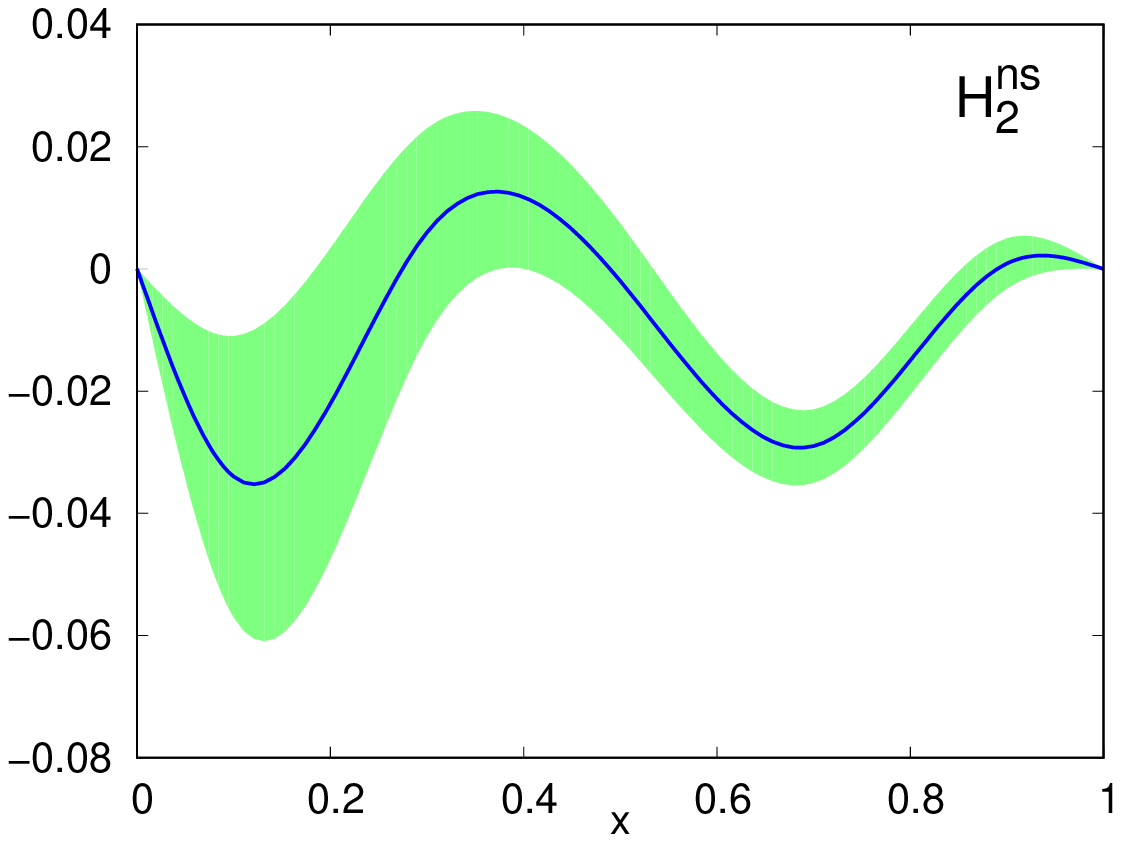}~~~~~\includegraphics[width=0.5\textwidth]{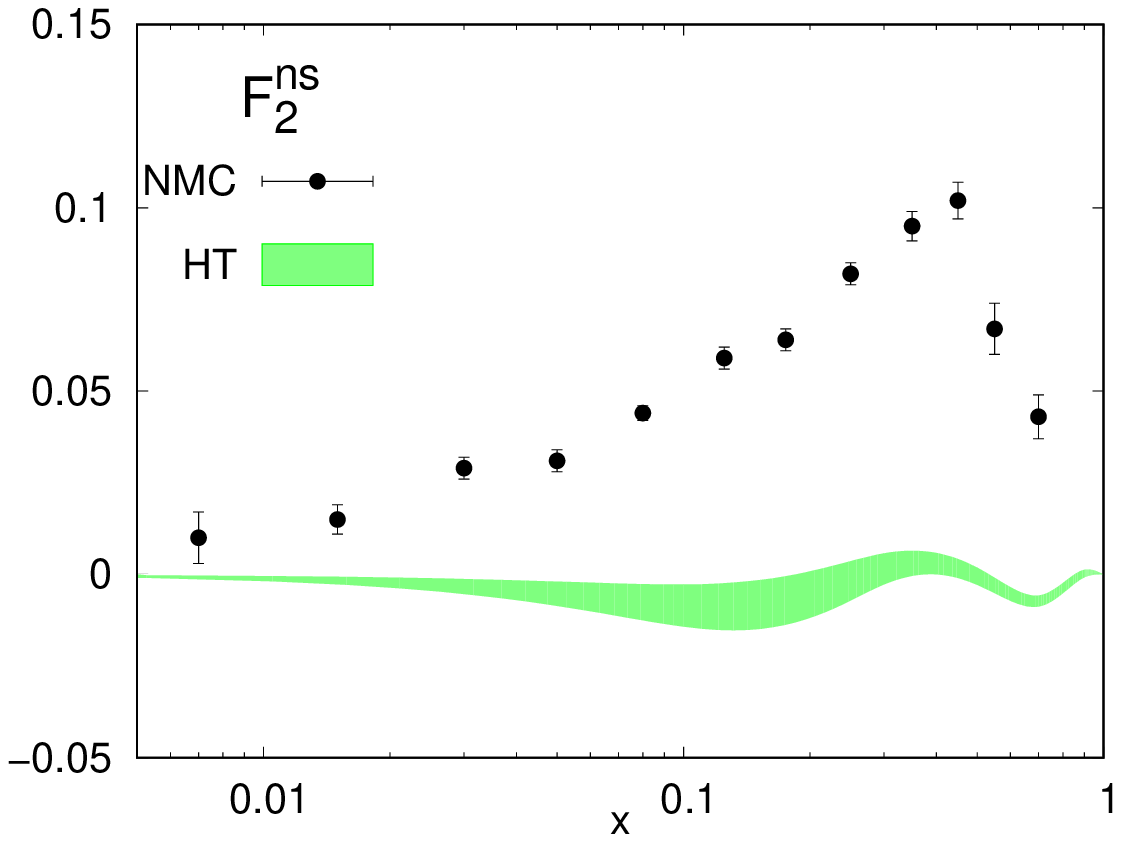}
\caption{\footnotesize \label{fig1}
Left: the central values (solid line) and the error band for the coefficient of the twist-4
term $H_2^{\tau=4}(x)$, Eq.~(\ref{eq2.14}), obtained from the NNLO fit for the nonsinglet
structure function $F_2^{p-n}(x)$ \cite{Alekhin:2012ig}. Right: comparison of
the NMC data for $F_2^{p-n}(x)$
at $Q^2=4\,{\rm GeV}^2$ with the corresponding HT corrections.}
\end{figure}

In light of the results obtained in \cite{Alekhin:2012ig}, also the higher-twist
terms seem not to be much helpful in description of the large discrepancy between
the theoretical prediction of the GSR, Eq.~(\ref{eq2.6}), and the experimental value of Eq.~(\ref{eq2.8}).
The authors of \cite{Alekhin:2012ig}, basing on the DIS world data, fitted in the NNLO
analysis the twist-4 coefficient $H_2^{\tau=4}(x)$ for the nonsinglet function
$F_2^{p-n}(x)$ and found the HT corrections marginal in comparison with the leading twist (LT) terms,
\begin{equation}\label{eq2.14}
F_2^{p-n}(x, Q^2) = F_2^p(x, Q^2) - F_2^n(x, Q^2)  = \left [F_2^{p-n}(x,Q^2)\right ]^{{\rm LT}} +
\frac{H_2^{\tau=4}(x)}{Q^2}\, .
\end{equation}
In Fig.~\ref{fig1}, we plot the coefficient of the twist-4 term $H_2^{\tau=4}(x)$, Eq.~(\ref{eq2.14}),
for the nonsinglet structure function $F_2^{p-n}(x)$ obtained in \cite{Alekhin:2012ig} and compare
the corresponding HT corrections with the results of NMC for $F_2^{p-n}(x)$ at $Q^2=4\,{\rm GeV}^2$.\\
However, when applied to the Gottfried sum rule, these HT corrections though too small to be
responsible for the observed flavor asymmetry, are not marginal and can
accurately explain the relatively large discrepancy between the two central
values of the experimental results: NMC, Eq.~(\ref{eq2.9}), and E866, Eq.~(\ref{eq2.10b}).\\
Namely, the HT effects modify the original GSR giving the contribution on the
level of $-5.4\%$ at $Q^2=4\,{\rm GeV}^2$ (NMC) and $-0.4\%$ at $Q^2=54\,{\rm GeV}^2$ (E866)
of the sum $1/3$. Hence, the corresponding difference between the NMC and E866 results for
the flavor asymmetry of the light sea
\begin{equation}\label{eq2.15}
\Delta(Q^2)\equiv\int^1_{0} \left [ \bar{d}(x,Q^2) - \bar{u}(x,Q^2)\right ] dx \, ,
\end{equation}
implied by the HT effects at different scales of $Q^2$, is
\begin{equation}\label{eq2.15a}
\Delta^{HT}(Q^2=4\,{\rm GeV}^2) - \Delta^{HT}(Q^2=54\,{\rm GeV}^2) \approx 0.025\pm 0.022\, .
\end{equation}
This is in a very good agreement with the experimental data:
\begin{equation}\label{eq2.16}
\Delta_{NMC} - \Delta_{E866} \approx 0.030\, .
\end{equation}
Taking into account also the perturbative QCD radiative corrections of Eq.~(\ref{eq2.12}),
we arrive at the value even closer to the data:  
\begin{equation}\label{eq2.17}
\Delta^{Rad+HT}(Q^2=4\,{\rm GeV}^2) - \Delta^{Rad+HT}(Q^2=54\,{\rm GeV}^2) \approx 0.027\, .
\end{equation}
It is seen that the $Q^2$-dependence of the GSR can resolve a discrepancy between the
flavor asymmetry of the light sea in the nucleon measured in different experiments.
Similar suggestion was made by the authors of Ref. \cite{Szczurek:1999wp}.
 
We have found that the QCD-improved parton model including the NNLO
radiative corrections and also the twist-4 contributions predicts for the Gottfried
sum rule at $Q^2=4\,{\rm GeV}^2$
\begin{equation}\label{eq2.18}
S_G(Q^2=4\,{\rm GeV}^2) = \frac{1}{3}\quad\underbrace{-0.0015}_{{\rm pQCD}}\quad
\underbrace{-0.0181}_{{\rm HT}} \approx \frac{1}{3} - 0.02\, .
\end{equation}
This means that the large deficit of the GSR observed in the experiments ($S_G\approx 1/3-0.1$)
comes from another sources than perturbative mechanisms and HT effects.

\subsection{Low-$x$ contribution}
\label{sec:sec2.3}
\begin{figure}[ht]
\centering
\includegraphics[width=0.5\textwidth]{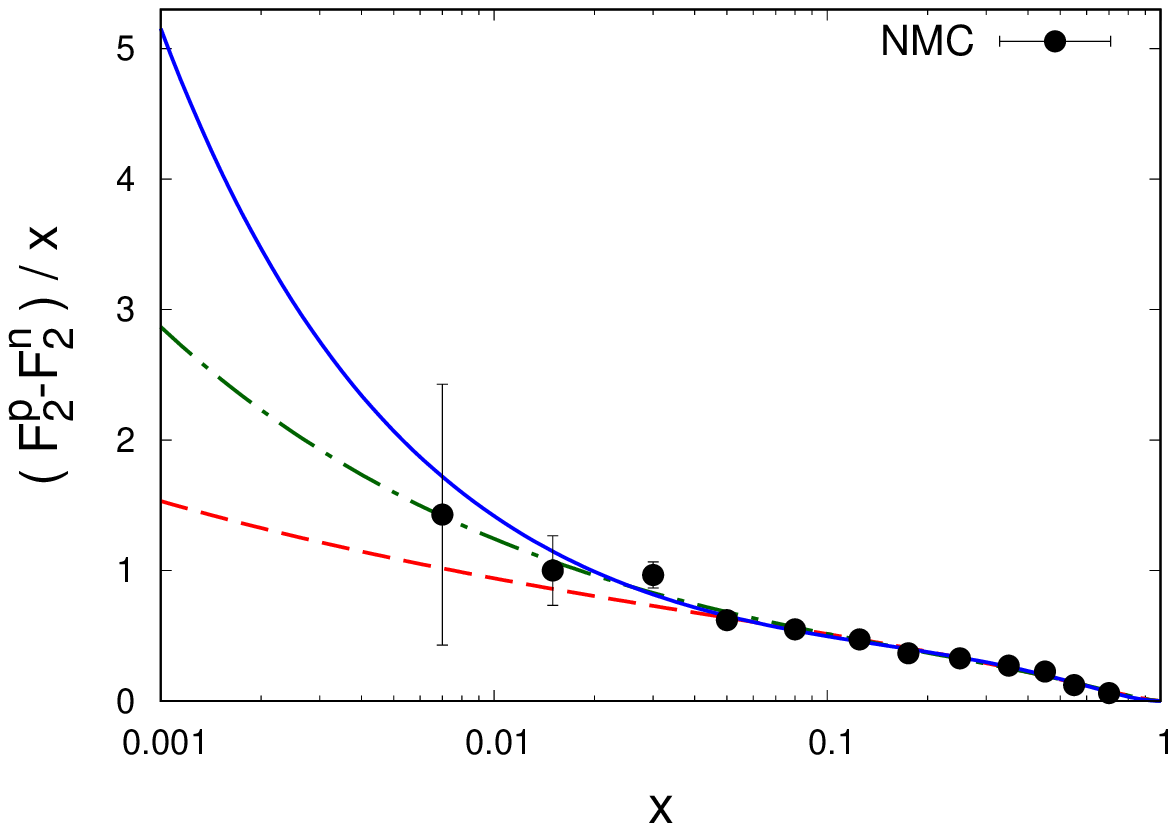}~~~~~\includegraphics[width=0.5\textwidth]{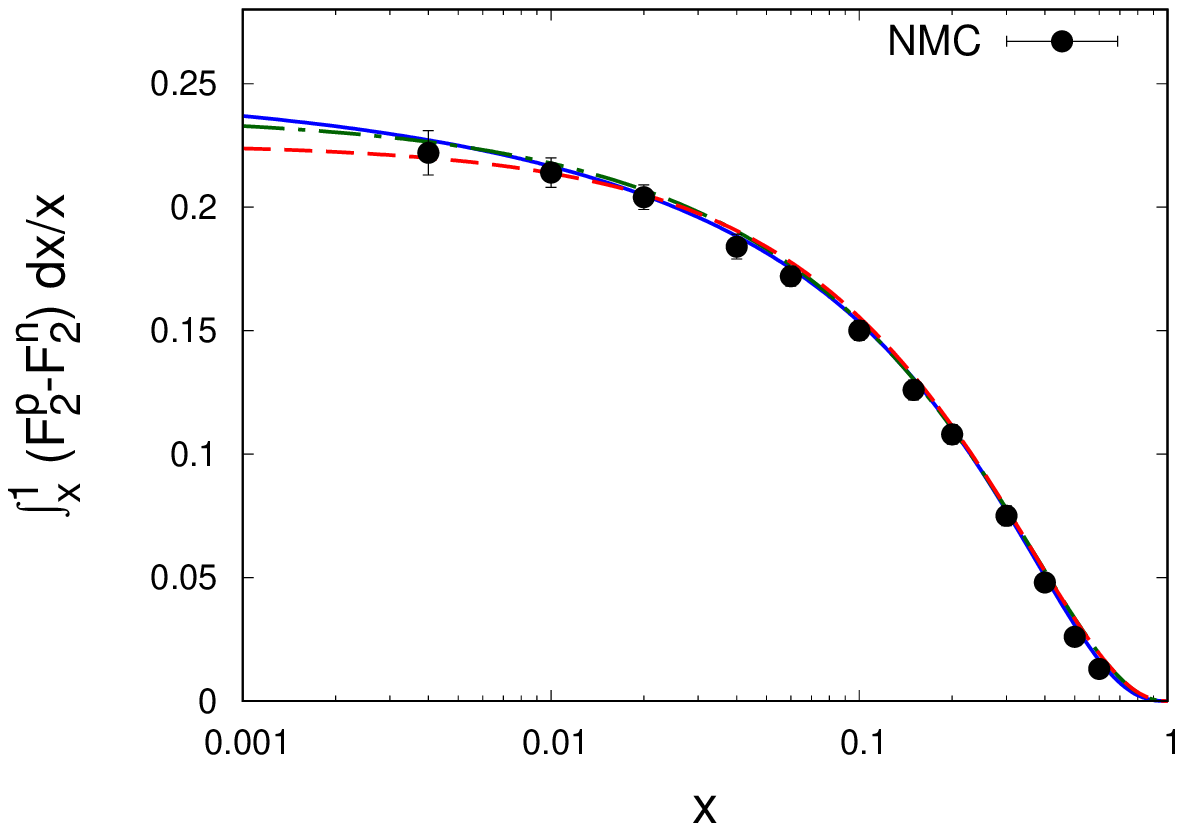}
\caption{\footnotesize \label{fig2}
Left: possible parametrizations of $F_2^{ns}/x$ reflecting different small-$x$
behavior $\sim x^a$: $a\,=\,-0.2$ (dashed), $-0.4$ (dash-dotted) and $-0.6$
(solid) compared to the NMC data \cite{Arneodo:1994sh}.
Right: the truncated Gottfried sum rule $\int_x^1 F_2^{ns}(x)\, dx/x$, respectively.}
\end{figure}

Experimental verification of the most sum rules faces the difficulty that in any realistic
experiment one cannot reach arbitrarily small values of the Bjorken $x$.
This is a serious obstacle also in the determination of the Gottfried sum rule which
involves the first Mellin moment, \ie integral of the nonsinglet structure
function $F_2^{ns}$ over the whole range of $x$:
$0\leqslant x\leqslant 1$. The lack of low-$x$ data with good accuracy makes
reasonable the idea that a significant contribution to the integral of the GSR
can come just from the small-$x$ region. We illustrate this in Fig.~\ref{fig2}
where we show different low-$x$ behaviors of $F_2^{ns}/x\sim x^a$ and the
corresponding truncated GSR, $\int_x^1 F_2^{ns}(x)\, dx/x$, together with
the NMC data \cite{Arneodo:1994sh}. We use three values for $a$: $-0.2$,
$-0.4$ and $-0.6$.
It is seen that the experimental uncertainties in the small-$x$ region are
too large to favor any of them.\\
 
The different $\sim x^a$ behaviors predict significant different very low-$x$
contributions to the GSR, $\int_0^{0.004} F_2^{ns}(x)\, dx/x$,
namely 0.006 for $a=-0.2$, 0.011 for $a=-0.4$ and 0.022 for $a=-0.6$.
This means that the very low-$x$ contributions to the GSR can vary from $1-7\,\%$ of
the sum $1/3$ and cannot resolve the GSR breaking problem. On the other hand,
the NMC data in the small-$x$ region confirm very well the expectations of the
theoretical studies on $F_2$ based on the Regge theory.
In the Regge approach, the small-$x$ behavior of $F_2^{ns}(x)$ is controlled by the
reggeon $A_ 2$ exchange \cite{Kwiecinski:1995rm}:
\begin{equation}\label{eq2.19}
F_2^{p-n}(x) = F_2^p(x) - F_2^n(x) \sim x^{1-\alpha_{A_2}}\, ,
\end{equation}
where $\alpha_{A_2}\approx 0.5$ is the $A_2$ reggeon intercept.
Taking into account the Regge predictions in the NMC data analysis, we can estimate
the small-$x$ contribution to the Gottfried sum $\int_0^{0.004} F_2^{ns}(x)\, dx/x$
as $4-7\,\%$ of the total value $1/3$.

\subsection{Nuclear shadowing}
\label{sec:sec2.4}  
Since there is no fixed target for the neutron, the deuteron is usually used for
measuring the neutron structure function $F_2^n$. The same method was used
by the NMC for determination of the Gottfried sum rule.
In order to obtain the difference of the structure functions $F_2^p-F_2^n$ of
free nucleons which enters into the GSR, the extracted $F_2^n$ from the deuteron
data has to be corrected by the shadowing effects:
\begin{equation}\label{eq2.20}
F_2^d = \frac{1}{2}\, \left( F_2^p + F_2^n\right) - \delta F_2^d\, ,
\end{equation}
where $\delta F_2^d\geqslant 0$.

\begin{figure}[ht]
\centering
\includegraphics[width=0.5\textwidth]{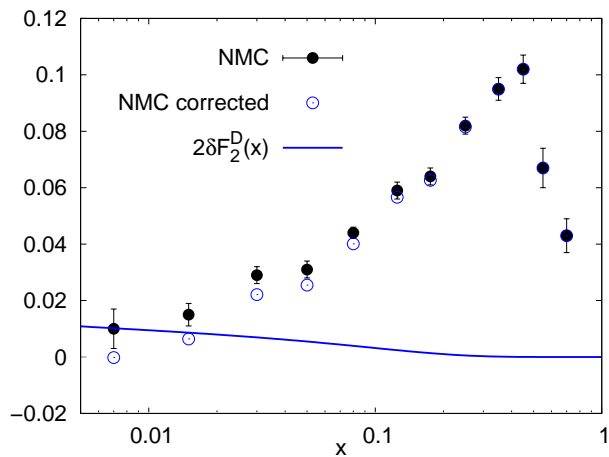}
\caption{\footnotesize \label{fig3}
The nuclear shadowing corrected NMC data for $F_2^{ns}$ (open circles) calculated
in \cite{Badelek:1994qg}. Solid: fit to the shadowing contribution to the
deuteron structure function, $2\delta F_2^d$.}
\end{figure}

The shadowing effects in the deuteron were investigated in many works
(for review, see, e.g., \cite{Kumano:1997cy}) providing the small negative
correction to the sum. Thus, the shadowing leads to smaller value of
$S_G(Q^2)$ than that determined experimentally assuming no shadowing,
and the GSR violation is even magnified.
The nuclear shadowing which is dominated by the vector-meson-dominance (VMD)
mechanism is non-negligible in the region of $x\leqslant 0.1$ and for low-
and moderate $Q^2$ relevant for the NMC measurements and has to be
taken into account in the data analysis \cite{Badelek:1994qg}.
This leads to the following expression for the difference between the proton
and neutron structure functions in the integrand of the Gottfried sum,
Eq.~(\ref{eq2.6}):
\begin{equation}\label{eq2.21}
F_2^p(x,Q^2) - F_2^n(x,Q^2) = \left ( F_2^p(x,Q^2) - F_2^n(x,Q^2) \right )_{NMC}
-2\,\delta F_2^d (x,Q^2)\, , 
\end{equation}
where $(F_2^p(x,Q^2) - F_2^n(x,Q^2))_{NMC}$ obtained by NMC is related to the measured
$F_2^d$ and $F_2^n/F_2^p$ via
\begin{equation}\label{eq2.22}
\left ( F_2^p(x,Q^2) - F_2^n(x,Q^2) \right )_{NMC} = 2\, F_2^d\;\frac{1-\left
(\frac{F_2^n}{F_2^p}\right )_{NMC}}{1+\left (\frac{F_2^n}{F_2^p}\right )_{NMC}}
\end{equation} 
Using the results of \cite{Badelek:1994qg} for $\delta F_2^d$, in Fig.~\ref{fig3},
we compare the NMC data with the corrected one by the nuclear shadowing effect.
We find that the negative shadowing correction to the experimental result for the Gottfried
sum, $S_G(0.004,\,0.8,\,Q^2) = 0.221$, is $0.0265$ ($\approx 12\%$).   

\section{TMM method for determination of sum rules}
\label{sec:sec3}
Here, we briefly present an effective method which allows one to determine
any sum rule value from the experimental data in the available restricted kinematic
range of the Bjorken variable $x$. The method was elaborated in
\cite{Kotlorz:2017wpu, Strozik-Kotlorz:2017gwn} for the Bjorken sum rule,
and successfully applied to the experimental data at COMPASS, SLAC and JLab
\cite{Kotlorz:2017wpu, Strozik-Kotlorz:2017gwn, Kotlorz:2018bxp, D.Kotlorz:2019oyu}.\\

The main philosophy of the method presented in \cite{Kotlorz:2017wpu} is
construction of a special truncated sum $\Gamma$ which approaches the limit of the sum
rule value more quickly, i.e. for larger $x$, than the ordinary sum.
In other words, the use of $\Gamma$ ``mimics'' the extension of the experimental
kinematic region of $x$ to the lower values.\\
Below, we give useful formulas for determination of the sum rule value in
the TMM approach which in the next section we shall apply to the GSR. The
details on theoretical aspects of the $\Gamma$ construction and the description of different
approximations of the TMM method can be found in \cite{Kotlorz:2017wpu}.\\

Determination of the sum rules involves the integrals of the parton density
or structure function $f(x,Q^2)$ over the whole range $(0,1)$ of $x$:
\begin{equation}\label{eq3.1}
S(0,\, 1) = \int^{1}_{0} f(x)\, dx\, ,
\end{equation}
where for clarity we omit the $Q^2$ dependence.
The experimental measurements provide data on $f(x)$ only in the limited range of $x$:
$0<x_{min}\equiv x_1<x_2<\cdots <x_{max}\equiv x_N<1$,
where $x_{min} \equiv Q^2_\text{min}/(2(Pq)_\text{max}>0)$.
Thus, in fact, the experiment gives information on the truncated sum 
\begin{equation}\label{eq3.2}
S(x_{min},\, x_{max}) = \int^{x_{max}}_{x_{min}} f(x)\, dx\, .
\end{equation}
The truncation at the upper limit $x_{max}$ is less important in comparison
to the low-$x$ limit $x_{min}$ because of the rapid decrease of the parton densities
and structure functions as $x\rightarrow 1$.\\

In particular, if we define $n$th truncated moment of the structure function $f(x,Q^2)$ as
\begin{equation}\label{eq2.11}
M_n(x_{min},\, x_{max},\, Q^2) = \int^{x_{max}}_{x_{min}} x^{n-2} f(x,Q^2)\, dx \,,
\end{equation}
the Gottfried sum rule $S_G(Q^2)$ is the first moment $M_1(0,\,1,\,Q^2)$ of the
nonsinglet function $F_2^p(x,\,Q^2) - F_2^n(x,\,Q^2)$.

\begin{figure}[h]
\centering
\includegraphics[width=0.5\textwidth]{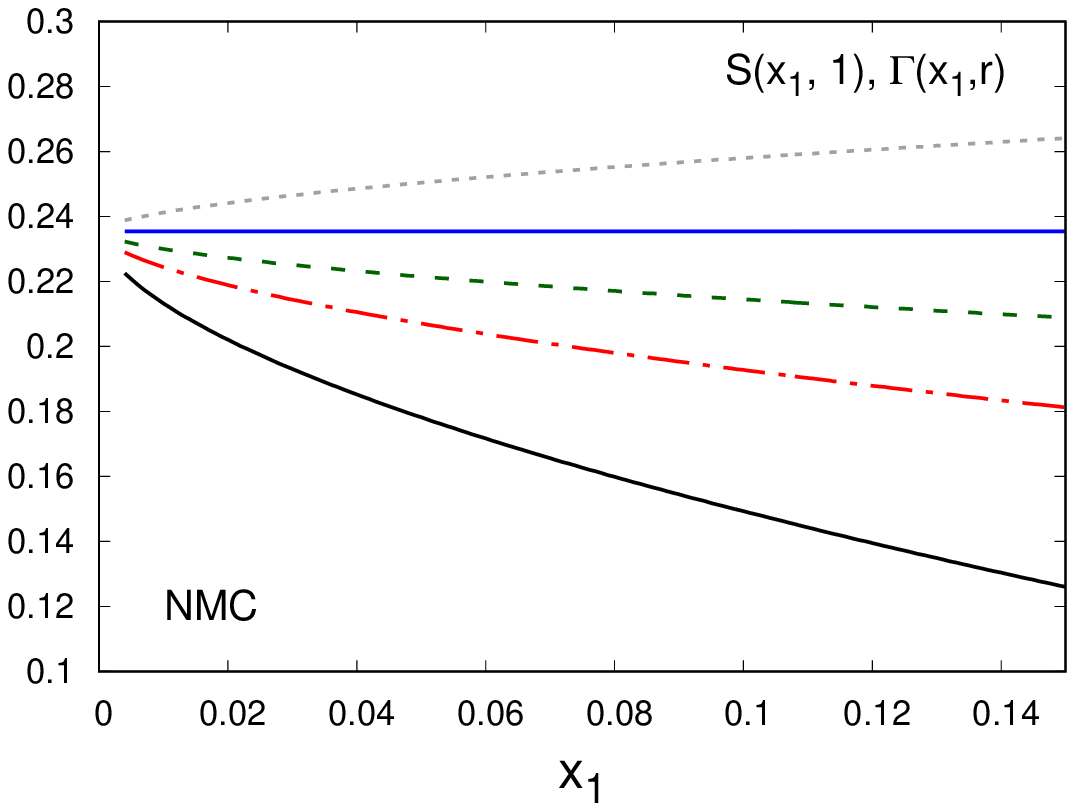}~~~~~\includegraphics[width=0.5\textwidth]{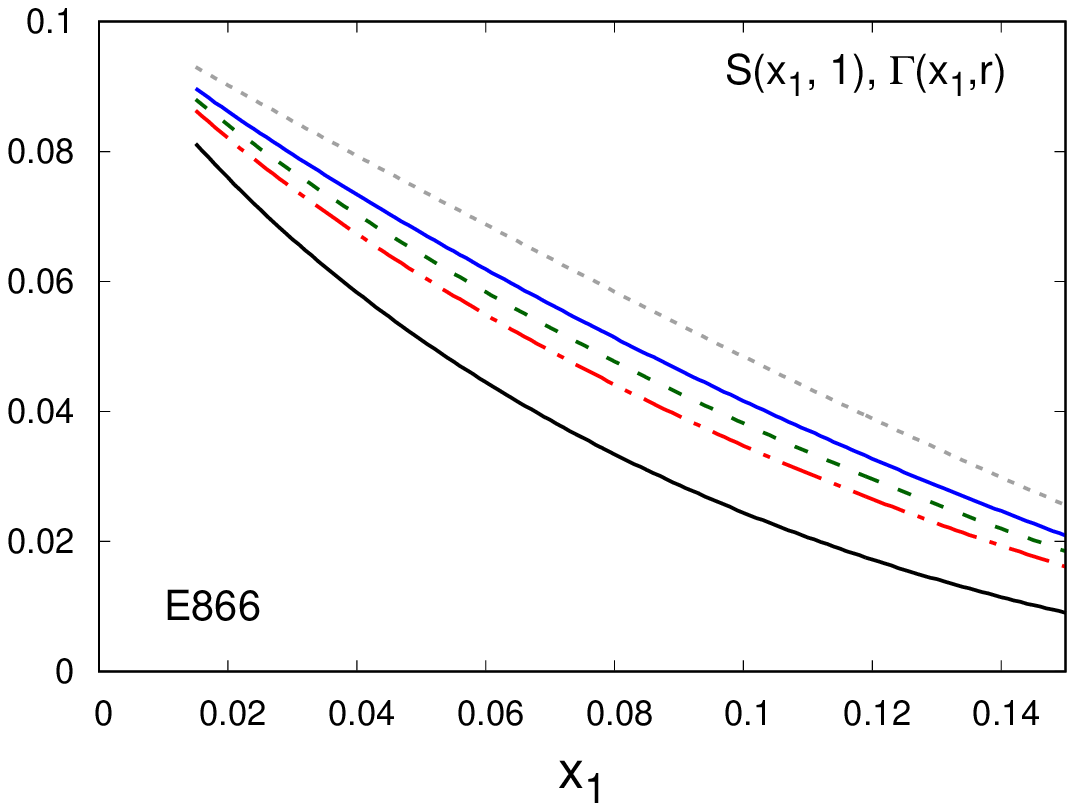}
\caption{\footnotesize \label{fig4}
$S(x_1,1)$, Eq.~(\ref{eq3.2}), (black solid) and $\Gamma (x_1,\, r)$, Eq.~(\ref{eq3.3}),
for different values of $A$. Upper (blue) solid line corresponds to $A=A_0$
(left panel) and $A=A_I$ (right panel), see description in the text.}
\end{figure}

The special sum $\Gamma$ is constructed based on the ordinary sum $S$ in
the following way:
\begin{eqnarray}
\Gamma (x_1,\, r) &=& S(x_1, 1) + A\,\int_{x_{1}}^{x_1/r} f(x)\,dx\,,\label{eq3.3}\\
\end{eqnarray}
where $x_1$ is the smallest value of $x$ accessible in the experiment and $A$,
and $r$ are parameters calculated from the data. 
In the limit $x_{1} \to 0$, $\Gamma (x_1,\, r)$ is equal to $S(x_1, 1)$ providing the sum rule
value $S(0,1)$, Eq.~(\ref{eq3.1}), whereas for $x>0$ $\Gamma (x_1,\, r)$
approaches $S(0,1)$ much earlier than $S(x_1,1)$ itself.
This is illustrated in Fig.~\ref{fig4} where we compare $S(x_1, 1)$ to a bunch of $\Gamma (x_1,\, r)$,
Eq.~(\ref{eq3.3}), plotted for different values of $A$. We use smooth fits to the NMC and E866 data
setting the ratio of two experimental points $r=x_1/x_2$ equal to $0.5$ for NMC and $0.7$
for E866, respectively. Here, we would like to emphasize that in our analysis we shall use
the auxiliary fit only for determination of $A_I$ while the rest of calculations will be
performed with use of the pure data $f(x)$ from the measured $x$-region.  

In our approach, as described in \cite{Kotlorz:2017wpu}, we utilize the quasi-linear
regime of $\Gamma (x_1,\, r)$ which starts already for $x$ significantly larger
than the smallest experimental value of $x_{1}$.
This ensures the applicability of the first (or even zero as in the NMC case)
order approximation for estimation of the value of $S(0,1)$ with help of $\Gamma (x_1,\, r)$.
Thus, requiring the second derivative to vanished, $\Gamma'' (x_1,\, r)=0$,
we obtain $A_I$ and the sum rule value can be determined very effectively in the
first order of Taylor expansion:
\begin{eqnarray}
S(0,1) = \Gamma (0,\, r) \approx \Gamma (x_1,\, r) - x_{1}\,\Gamma'(x_{1},\, r) 
&=& \Gamma (x_1,\, r) + (A_I+1)\,x_{1}\, f(x_{1}) - A_I\,\frac{x_{1}}{r}f(x_{1}/r) \, , \label{eq3.5} \\
A_I &=& \left[ \frac{1}{r^2}\,\frac{f'(x_{1}/r)}{f'(x_{1})}-1\right]^{-1}\, ,
\label{eq3.6}
\end{eqnarray}
where $\Gamma (x_1,\, r)$ is given by Eq.~(\ref{eq3.3}) and $f'(x)$ denotes
the first order derivative with respect of $x$.\\

In a special case, where the small-$x$ experimental data can be well
described by a simple form $f(x)=Nx^a$, we have
\begin{equation}\label{eq3.7}
r^n\frac{f^{(n-1)}(x_1)}{f^{(n-1)}(x_1/r)} = r\frac{f(x_1)}{f(x_1/r)}
\end{equation}
and all derivatives $\Gamma^{(n)} (x_1,\, r)$ vanish for the same $A=A_0$,
\begin{equation}\label{eq3.8}
A_0 = \left[ \frac{1}{r}\,\frac{f(x_{1}/r)}{f(x_{1})}-1\right]^{-1}\, .
\end{equation}
Hence, we arrive at the zero order approximation for the sum rule value
which reads
\begin{equation}\label{eq3.9}
S(0,1) = \Gamma (x_1,\, r) = S(x_1, 1) + A_0\,\int_{x_{1}}^{x_1/r} f(x)\,dx\, .
\end{equation}

The method of estimation of the sum rule value based on the special
truncated sum $\Gamma$ is effective for different small-$x$ behavior of the
function $f\sim x^a$, also for $a<0$, as in the case of the Gottfried sum rule.

\section{Data analysis}
\label{sec:sec4}
Below we present our numerical results for the Gottfried sum rule value
$S(0,1)$ based on the experimental NMC~\cite{Arneodo:1994sh} and E866
\cite{Towell:2001nh} data following the approach described in the previous section.

\subsection{NMC}
\label{sec:sec4.1}
The violation of the GSR was first observed by the New
Muon Collaboration at CERN in 1991 \cite{Amaudruz:1991at}.
NMC measured the cross section ratio for deep inelastic scattering of muons
from hydrogen and deuterium targets in the kinematic range extended to the
low-$x$ region, $0.004\leqslant x\leqslant 0.8$. The difference of the structure
functions was calculated by Eq.~(\ref{eq2.22}) and the ratio
$F_2^n/F_2^p = 2F_2^d/F_2^p - 1$ was determined by the NMC experiment, where
the deuteron structure function $F_2^d$ was taken from a fit to various
experimental data. The results were obtained by interpolation or
extrapolation to $Q^2 = 4{\rm\,GeV}^2$.
In the reanalyzed data \cite{Arneodo:1994sh}, which are under study
in this section, NMC used their own data for $F_2^d$ and revised
$F_2^n/F_2^p$ ratios.
  
The NMC data for $F_2^{p-n}$ which form the GSR,
\begin{equation}\label{eq4.1}
S_G(0,1) = \int^1_{0} \left [F_2^p(x,Q^2) - F_2^n(x,Q^2)\right ] {dx\over x}\, ,
\end{equation} 
can be for $x\leqslant 0.4$ well described by the fit function $\sim 0.2\,x^{0.6}$
\cite{Arneodo:1994sh}, 
which agrees with theoretical prediction of the Regge-like behavior, 
Eq.~(\ref{eq2.19}), \cite{Kwiecinski:1995rm}. The corresponding truncated function
$\Gamma (x_1,\, r)$ saturates to the constant $S_G(0,1)$ already at large $x$
(see upper solid line in the left panel of Fig.~4) and we can estimate
$S_G(0,1)$ using the zero order formulas, Eqs.~(\ref{eq3.8}) and (\ref{eq3.9}),
which take the form
\begin{eqnarray}
S_G(0,1) &=& S_G(x_{1},1)^{\rm{NMC}} + A_0\,\int_{x_{1}}^{x_1/r} F_2^{p-n}(x,Q^2) \frac{dx}{x}\, ,
\label{eq4.2} \\
A_0 &=& \left[\frac{F_2^{p-n}(x_{1}/r,Q^2)}{F_2^{p-n}(x_{1},Q^2)}-1\right]^{-1}.
\label{eq4.3}
\end{eqnarray}
All quantities in Eqs.~(\ref{eq4.2}) and (\ref{eq4.3}) are directly provided
by the data or can be calculated from the data without necessity to use of any fit function.
Namely, $S_G(x_{1},1)^{\rm{NMC}}$ is the contribution to the GSR
from the measured region of $x$ together with the correction for $x>0.8$,
$x_1$ denotes the smallest $\langle x\rangle$ in the analysis, and $r=x_1/x_k$ is a ratio of
two experimental points where $x_k>x_1$.
The integral in Eq.~(\ref{eq4.2}) can be calculated as a sum of the partial
experimental contributions, respectively:
\begin{equation}\label{eq4.2a}
\int_{x_i}^{x_{i+1}} F_2^{p-n}(x)\,\frac{dx}{x} =
\frac{x_{i+1}-x_i}{\langle x_i\rangle}\,F_2^{p-n}(\langle x_i\rangle)\, . 
\end{equation}

In Table~\ref{tab1} we show our estimations for $S_G(0,1)$ obtained for
two values of $x_1$: $0.007$ and $0.015$ and corresponding $r$ and $A_0$, up
to $x_k\approx 0.4$. The experimental value of $S_G(0,1)$, where the small-$x$
contribution from the region $x<0.004$ is determined from the fit
described in \cite{Arneodo:1994sh}, is displayed in the last row.
\begin{table}[h]
\centering
\caption{Estimations of the Gottfried sum rule value $S_G(0,1)$ obtained in
the zero order approximation, Eqs.~(\ref{eq4.2}) and (\ref{eq4.3}), for
two values of $x_1$: $0.007$ and $0.015$ based on the NMC data
at $Q^2=4$~\rm{GeV}$^2$ \cite{Arneodo:1994sh}. The ratio $r=x_1/x_{1+i}$,
where $i=1,\,2,\, ...\, 8$.
The experimental value of $S_G(0,1)$ is displayed in the last row.}
\label{tab1}
\begin{tabular}{cccccc}\hline
\multicolumn{3}{c}{$x_1=0.007$} & \multicolumn{3}{c}{$~~~~~~x_1=0.015$} \\
$r$ & $~~~A_0$ & $~~~S_G(0,1)$ & $~~~~~~r$ & $~~~A_0$ & $~~~S_G(0,1)$ \\
\hline\hline
$0.47$ & $~~~2.0$  & $~~~0.236$ & $~~~~~~0.50$ & $~~~1.07$ & $~~~0.223$ \\
$0.23$ & $~~~0.53$ & $~~~0.230$ & $~~~~~~0.30$ & $~~~0.94$ & $~~~0.236$ \\
$0.14$ & $~~~0.48$ & $~~~0.236$ & $~~~~~~0.19$ & $~~~0.52$ & $~~~0.232$ \\
$0.09$ & $~~~0.29$ & $~~~0.234$ & $~~~~~~0.12$ & $~~~0.34$ & $~~~0.232$ \\
$0.06$ & $~~~0.20$ & $~~~0.233$ & $~~~~~~0.09$ & $~~~0.31$ & $~~~0.236$ \\
$0.04$ & $~~~0.19$ & $~~~0.236$ & $~~~~~~0.06$ & $~~~0.22$ & $~~~0.234$ \\
$0.03$ & $~~~0.14$ & $~~~0.235$ & $~~~~~~0.04$ & $~~~0.19$ & $~~~0.235$ \\
$0.02$ & $~~~0.12$ & $~~~0.235$ & $~~~~~~0.03$ & $~~~0.17$ & $~~~0.237$ \\
\hline
\multicolumn{6}{c}{ EXP. NMC $S_G(0,1)=0.235 \pm 0.026$} \\ \hline
\end{tabular}
\end{table}

We obtain $S_G(0,1)=0.234 \pm 0.003_{\,\rm SD}$.
The low-$x$ contribution $S_G(0, 0.004)=0.012 \pm 0.004$.
Here, we estimate the error for the low-$x$ contribution as
an average deviation for the composed errors of $S_G(0, 0.004)$ calculated
for the data sets summarized in Table~\ref{tab1}.
Hence, we find finally $S_G(0,1)=0.234 \pm 0.022$.
The obtained result is in a good agreement with the value provided by NMC.
This means that in the case of the NMC data, the Gottfried sum can be determined
very effectively already in the zero order approximation of the TMM approach.\\

\begin{figure}[h]
\centering
\includegraphics[width=0.5\textwidth]{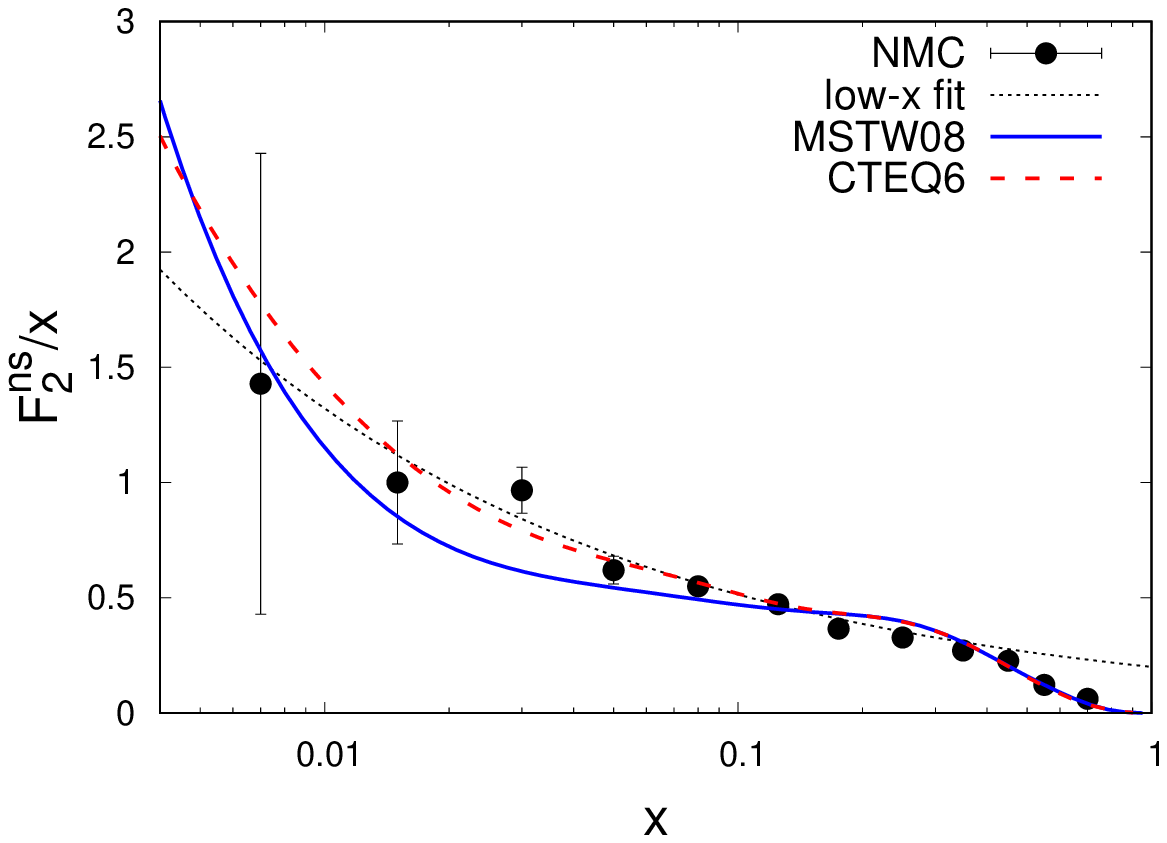}~~~~~\includegraphics[width=0.5\textwidth]{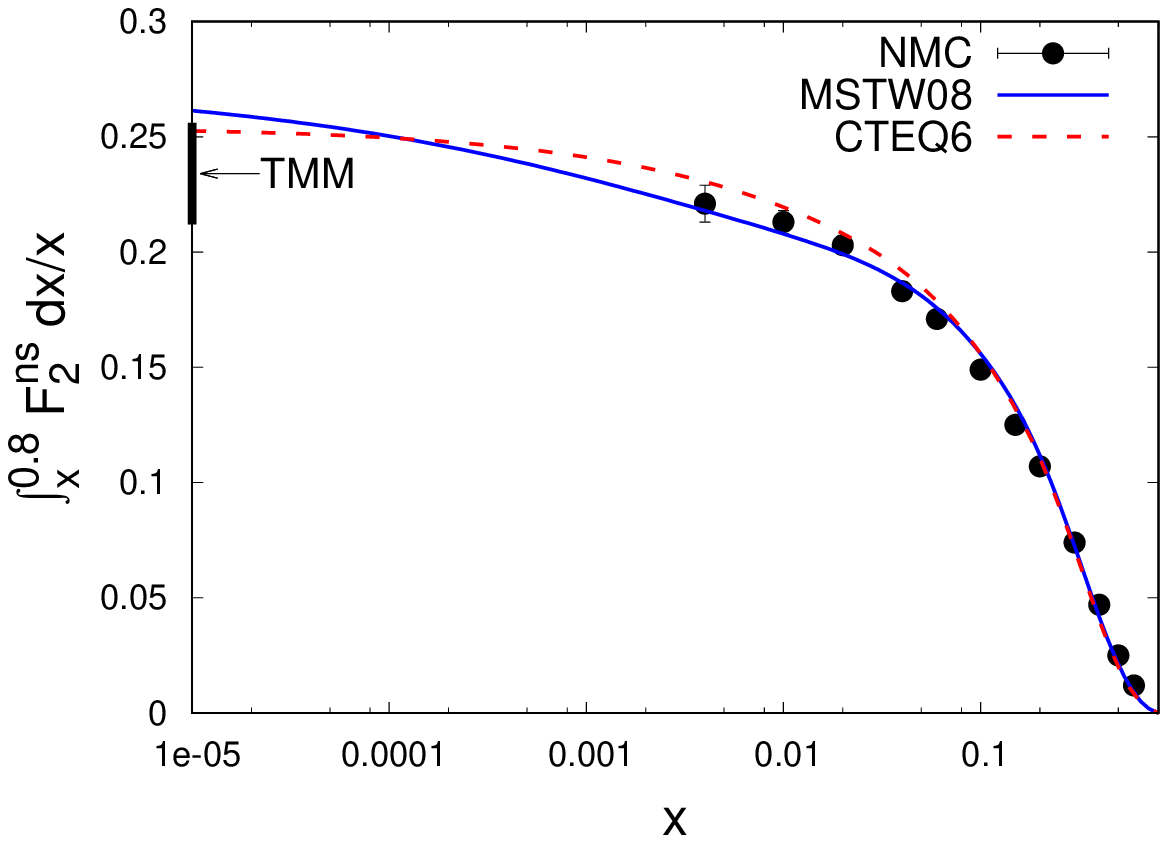}
\caption{\footnotesize \label{fig5}
$F_2^{ns}(x,Q^2)/x$ (left) and $\int_x^{0.8} F_2^{ns}(x,Q^2)\,dx/x$ (right)
at $Q^2=4\,{\rm GeV^2}$.
A comparison of the NMC data to CTEQ6 \cite{Pumplin:2002vw} and MSTW08 \cite{Martin:2009iq}
predictions and also to the TMM estimation for the Gottfried sum rule.
The low-$x$ fit function $f_{\rm fit}^{\rm NMC}$ in the left panel has the form Eq.~(\ref{fit_NMC}).}
\end{figure}

In Fig.~\ref{fig5}, we compare the NMC data for
$F_2^{ns}/x$ and $\int_x^{0.8} F_2^{ns} dx/x$ with the predictions of
two parametrizations based on the global PDFs fit, CTEQ6 \cite{Pumplin:2002vw}
and MSTW08 \cite{Martin:2009iq}, and to our TMM estimation for the GSR.
In the left panel we plot also the low-$x$ fit function $f_{\rm fit}^{\rm NMC}$
for illustration of the regular Regge behavior of the NMC data up to $x\approx 0.4$.
We find
\begin{equation}
\frac{1}{x}\left [F_2^p(x,Q^2)-F_2^n(x,Q^2)\right ]_{\rm NMC}\approx f_{\rm fit}^{\rm NMC}
= ax^b,\quad\quad a = 0.198\pm 0.006, \quad b = -0.410\pm 0.012, 
\label{fit_NMC}
\end{equation}
which is consistent with the form provided in \cite{Arneodo:1994sh}.

In Table ~\ref{tab2}, we collect the contributions to the Gottfried sum rule,
$\int F_2^{ns}(x,Q^2)\, dx/x$, obtained for different $x$ ranges at
$Q^2=4\, {\rm GeV}^2$. A comparison of the NMC data to TMM, CTEQ6 and MSTW08 predictions
is shown.\\
\begin{table}[h]
\centering
\caption{The contributions to the Gottfried sum rule, $\int F_2^{ns}(x,Q^2)\, dx/x$,
integrated over different $x$ ranges at $Q^2=4\, {\rm GeV}^2$. Compared are
the NMC data to TMM,
CTEQ6 \cite{Pumplin:2002vw} and MSTW08 \cite{Martin:2009iq} predictions.
$^{(*)}$ The result contains a fit to the unmeasured region of small-$x$.}
\label{tab2}
\begin{tabular}{ccccc}\hline
$~~~~~x$ range$~~~~~$ & $~~~$NMC$~~~$ & $~~~$TMM$~~~$  & $~~~$CTEQ6$~~~$ & $~~~$MSTW08$~~~$ \\
\hline\hline
$0<x<1$        & $0.235\pm 0.026^*$      & $0.234\pm 0.022$  & $0.255$   & $0.274$  \\
$0.004<x<0.8$  & $~~~0.221\pm 0.021~~~$  &                   & $0.231$   & $0.218$  \\
$0<x<0.004$    & $0.013\pm 0.005^*$      & $0.012\pm 0.004$  & $0.023$   & $0.055$  \\
$0.8<x<1$      & $0.001\pm 0.001$        &                   & $0.001$   & $0.001$  \\
\hline
\end{tabular}
\end{table}

The small-$x$ contribution to the GSR in the unmeasured region $x<0.004$
was determined by NMC with use of the fit to the data.
Our method, which is totally based on the experimental data in the measured
region of $x$, provides almost the same result. 
The CTEQ6 prediction is slightly above the NMC estimation for $S_G(0,0.004)$
while agreeing for the total GSR value and also for the contribution from the measured region.
In turn, the MSTW08 parametrization supports the experimental measurements but its predictions
for $S_G(0,0.004)$ and $S_G(0,1)$ are much larger than both the experimental and our
estimations.
A reason for this discrepancy is a small-$x$ behavior of $\bar{d} - \bar{u}$ assumed
by MSTW08 which implies a decrease of $\int (\bar{d} - \bar{u})\, dx$ and hence the increase of
$\int F_2^{ns}/x\, dx$ in this region in comparison to other global PDF fits.
We shall discuss it also in the next subsection which is devoted to the E866 experiment.   

\subsection{E866}
\label{sec:sec4.2}
Fermilab experiment E866 \cite{Towell:2001nh} was a fixed target experiment that has
measured the light sea quark asymmetry in the nucleon using Drell-Yan
process of di-muon production in 800 GeV proton interactions with hydrogen
and deuterium targets.
From the data, the ratio $\bar{d}/\bar{u}$ was determined over a wide range in Bjorken-$x$.
The obtained results confirmed previous measurements by E866/NuSea
\cite{Hawker:1998ty}, which were the first demonstration of a strong $x$-dependence
of the $\bar{d}/\bar{u}$ ratio, and extended them to lower-$x$.
To obtained the antiquark asymmetry $\bar{d}-\bar{u}$ and also the integrated
asymmetry $\int (\bar{d}-\bar{u})\, dx$, E866 used their data for $\bar{d}/\bar{u}$
and the PDF parametrization CTEQ5M \cite{Lai:1999wy} for $\bar{d}+\bar{u}$.
In order to estimate the contribution from the unmeasured region $0<x<0.015$,
MRST \cite{Martin:1998sq} and CTEQ5M fits were used. Moreover, it was assumed that the
contribution for $x>0.35$ was negligible.\\
In our TMM analysis, presented below, we use the experimental data only from
the measured region $0.015<x<0.35$. We compare our results with the E866 ones
and also with the predictions of the updated global parametrizations -- CTEQ6
\cite{Pumplin:2002vw} and MSTW08 \cite{Martin:2009iq}.\\

\begin{figure}[h]
\centering
\includegraphics[width=0.5\textwidth]{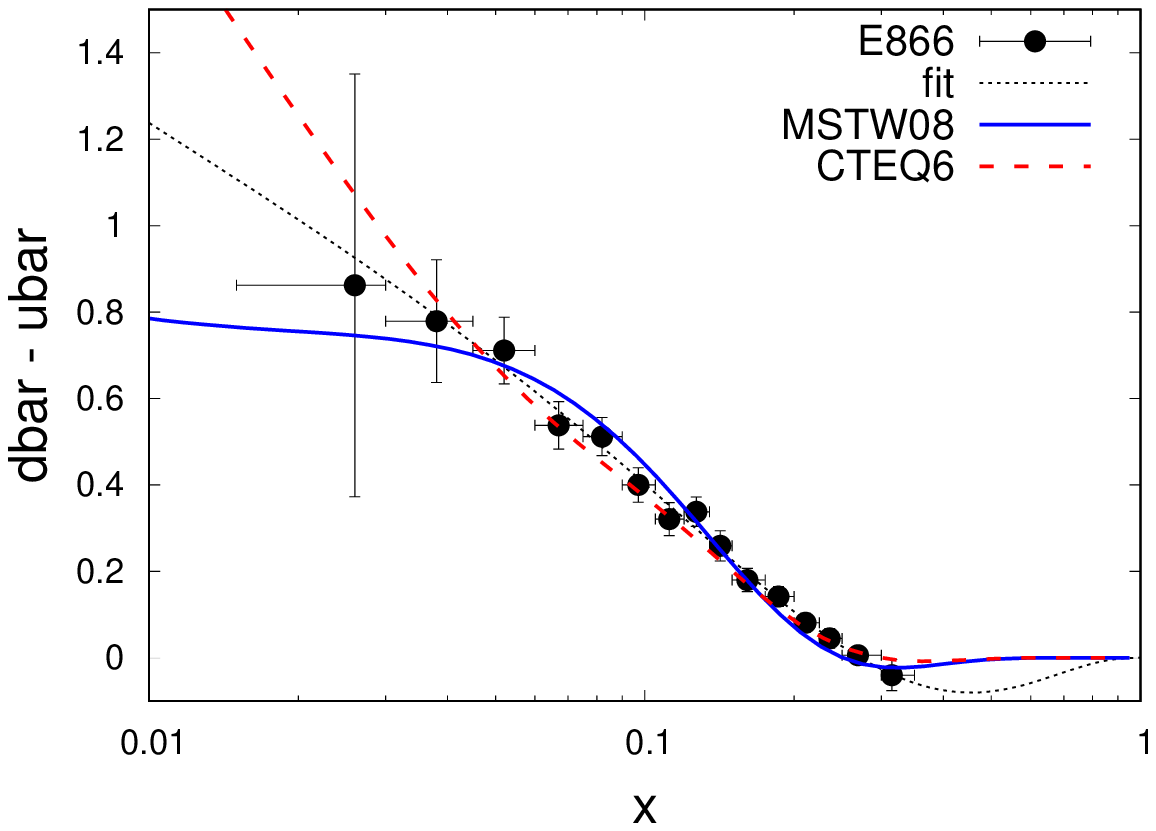}~~~~~\includegraphics[width=0.5\textwidth]{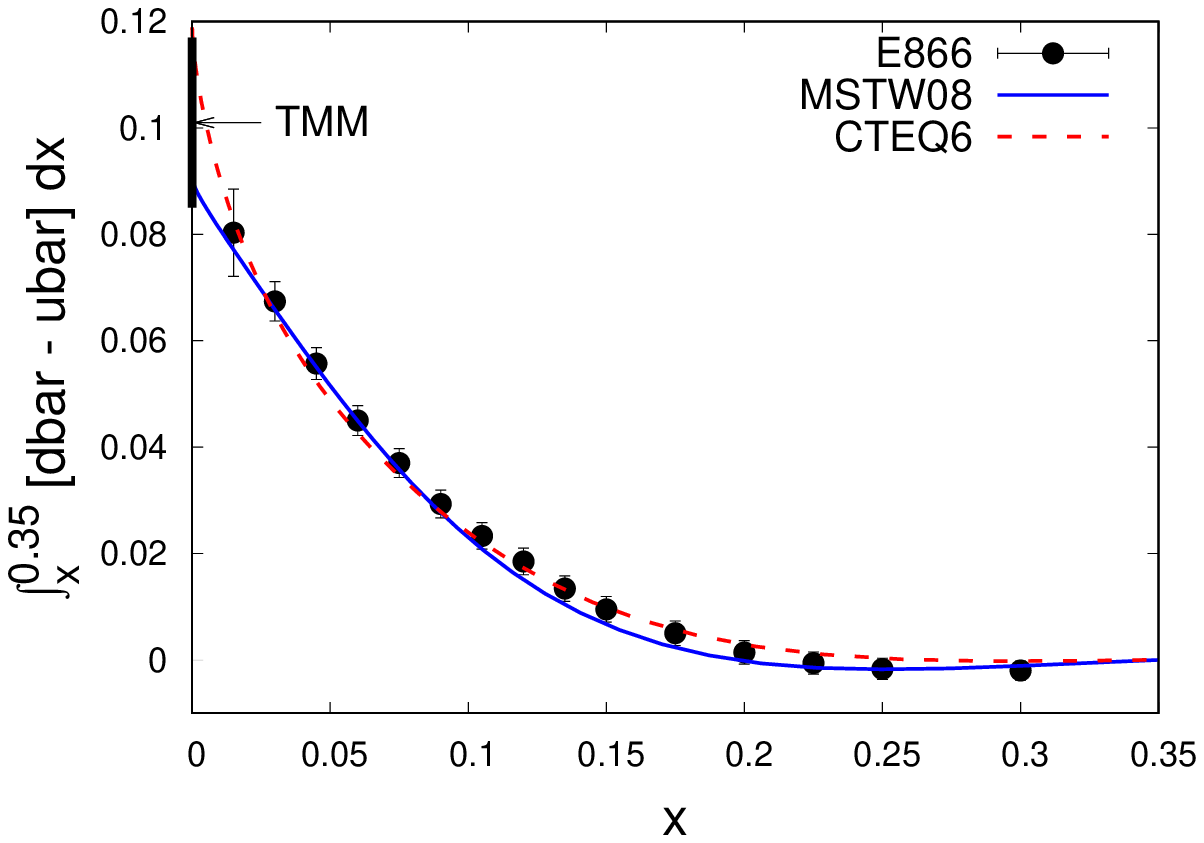}
\caption{\footnotesize \label{fig6}
$[\bar{d}-\bar{u}](x,Q^2)$ (left) and $\int_x^{0.35} [\bar{d}-\bar{u}](x,Q^2)\,dx$ (right)
at $Q^2=54\,{\rm GeV^2}$.
A comparison of the E866 data to CTEQ6 \cite{Pumplin:2002vw} and MSTW08 \cite{Martin:2009iq}
predictions and also to the TMM estimation for the integrated asymmetry of
the light sea quarks $\int_0^1 [\bar{d}-\bar{u}](x,Q^2)\,dx$.
The fit function $f_{\rm fit}^{\rm E866}$ in the left panel has the form Eq.~(\ref{fit_E866}).}
\end{figure}

To determine the light sea quark asymmetry from the E866 data,
\begin{equation}\label{eq4.4}
\Delta(0,1) = \int^1_{0} \left [\bar{d}(x,Q^2) - \bar{u}(x,Q^2)\right ] dx\, ,
\end{equation}
we apply the universal first order approximation of the special truncated sum
$\Gamma$ method given by Eqs.~(\ref{eq3.3})-(\ref{eq3.6}).
In the terms of the experimental data they read
\begin{eqnarray}
\Delta(0,1) = \Delta(x_{1},1)^{\rm{E866}} + &A_I&\,\int_{x_{1}}^{x_1/r}
[\bar{d}-\bar{u}](x)\, dx
 + (A_I+1)\,x_{1}\, [\bar{d}-\bar{u}](x_{1}) - A_I\,\frac{x_{1}}{r}[\bar{d}-\bar{u}](x_{1}/r) \, , \label{eq4.5} \\
&A_I& = \left[\frac{1}{r^2}\,\frac{f_{fit}'(x_{1}/r)}{f_{fit}'(x_{1})}-1\right]^{-1}\, ,
\label{eq4.6}
\end{eqnarray}
where the prime denotes a derivative with respect to $x$ of the fit function
$f_{\rm fit}^{\rm E866}$ to the E866 data on $\bar{d}-\bar{u}$:
\begin{eqnarray}
\left [\bar{d}(x,Q^2)-\bar{u}(x,Q^2)\right ]_{\rm E866}\approx f_{\rm fit}^{\rm E866} &=&
ax^b(1-x)^c(1+d\,x), \nonumber \\
a = 0.55\pm 0.02, \quad b = -0.19\pm 0.02, \quad c &=& 2.8\pm 0.3 , \quad d = -3.7\pm 0.1\, . 
\label{fit_E866}
\end{eqnarray}
The fit function, which we use only for calculation of $A_I$ in Eq.~(\ref{eq4.6}),
is shown as a dotted line in the left panel of Fig.~\ref{fig6}.
All other quantities in Eqs.~(\ref{eq4.5}) and (\ref{eq4.6}) are directly provided
by the data. Again, as in the NMC analysis, $x_1$ denotes the smallest $\langle x\rangle$,
and the ratio  $r=x_1/x_k$ is determined from the kinematics.
$\Delta(x_{1},1)^{\rm{E866}}$ denotes the contribution to the light sea
asymmetry from the measured region of $x$.\\
\begin{table}[h]
\centering
\caption{Estimations of the integrated quark asymmetry $\Delta(0,1)$ obtained in
the first order approximation, Eqs.~(\ref{eq4.5}) and (\ref{eq4.6}), for
two values of $x_1$: $0.026$ and $0.038$ based on the E866 data
at $Q^2=54$~\rm{GeV}$^2$, \cite{Towell:2001nh}. The ratio $r=x_1/x_{1+i}$
where $i=1,\,2,\,3,\, ...\, 8$.
The experimental value of $\Delta(0,1)$ is displayed in the last row.}
\label{tab3}
\begin{tabular}{cccccc}\hline
\multicolumn{3}{c}{$x_1=0.026$} & \multicolumn{3}{c}{$~~~~~~x_1=0.038$} \\
$r$ & $~~~A_I$ & $~~~\Delta(0,1)$ & $~~~~~~r$ & $~~~A_I$ & $~~~\Delta(0,1)$ \\
\hline\hline
$0.68$ & $~~~1.79$ & $~~~0.098$ & $~~~~~~0.73$ & $~~~2.19$ & $~~~0.097$ \\
$0.50$ & $~~~0.79$ & $~~~0.098$ & $~~~~~~0.57$ & $~~~1.03$ & $~~~0.105$ \\
$0.39$ & $~~~0.48$ & $~~~0.101$ & $~~~~~~0.46$ & $~~~0.67$ & $~~~0.101$ \\
$0.32$ & $~~~0.35$ & $~~~0.099$ & $~~~~~~0.39$ & $~~~0.50$ & $~~~0.103$ \\
$0.27$ & $~~~0.27$ & $~~~0.101$ & $~~~~~~0.34$ & $~~~0.40$ & $~~~0.104$ \\
$0.23$ & $~~~0.22$ & $~~~0.101$ & $~~~~~~0.30$ & $~~~0.33$ & $~~~0.101$ \\
$0.20$ & $~~~0.19$ & $~~~0.100$ & $~~~~~~0.27$ & $~~~0.29$ & $~~~0.103$ \\
$0.18$ & $~~~0.17$ & $~~~0.101$ & $~~~~~~0.24$ & $~~~0.26$ & $~~~0.105$ \\
\hline
\multicolumn{6}{c}{ EXP. E866 $\Delta(0,1)=0.118 \pm 0.012$} \\ \hline
\end{tabular}
\end{table}
In Table~\ref{tab3} we show our estimations for $\Delta(0,1)$ obtained for
two values of $x_1$: $0.026$ and $0.038$ and corresponding $r$ and $A_I$.
To minimize a possible error implied by the decrease of $r$, \cite{Kotlorz:2017wpu},
we proceed our analysis up to $x_k\approx 0.15$.\\
In Fig.~\ref{fig6}, we compare the E866 data for
$[\bar{d}-\bar{u}](x,Q^2)$ and \mbox{$\int_x^{0.35}[\bar{d}-\bar{u}](x,Q^2)\,dx$}
with the predictions of two parametrizations based on the global PDFs fits,
CTEQ6 \cite{Pumplin:2002vw} and MSTW08 \cite{Martin:2009iq}, and to the TMM
results for \mbox{$\int_0^1 [\bar{d}-\bar{u}](x,Q^2)\,dx$}.
In Table ~\ref{tab4}, we present a comparison of the light sea quark
asymmetry $\Delta$ integrated over different $x$ ranges for the E866 data,
TMM approach and CTEQ6 and MSTW08 predictions. The TMM results are shown
together with the average deviation of the composed errors calculated from
Eqs.~(\ref{eq4.5}) and (\ref{eq4.6}) for the set data from Table~\ref{tab3}.\\
\begin{table}[h]
\centering
\caption{Integrated light sea quark asymmetry $\Delta$ over different
$x$ ranges at $Q^2=54$ ${\rm GeV^2}$ obtained in the TMM approach.
A comparison to the E866 data and
CTEQ6 \cite{Pumplin:2002vw} and MSTW08 \cite{Martin:2009iq} predictions.
$^{(*)}$ The result contains a fit to the unmeasured region of small-$x$.}
\label{tab4}
\begin{tabular}{ccccc}\hline
$~~~~~x$ range$~~~~~$ & $~~~$E866$~~~$ & $~~~$TMM$~~~$ & $~~~$CTEQ6$~~~$ & $~~~$MSTW08$~~~$ \\
\hline\hline
$0<x<1$         & $0.118\pm 0.012^*$       & $0.101\pm 0.016$   & $0.119$  & $0.089$   \\
$0.015<x<0.35$  & $~~~0.0803\pm 0.011~~~$  &                    & $0.083$  & $0.077$   \\
$0<x<0.015$     & $0.038\pm 0.004^*$       & $0.021\pm 0.012$   & $0.037$  & $0.014$  \\
$0.35<x<1$      & $0$                      &   & $\!\!\!-0.001$ & $\!\!\!-0.002$ \\
\hline
\end{tabular}
\end{table}
The low-$x$ contribution $\Delta(0, 0.015)=0.021 \pm 0.012$ obtained in our
analysis is essentially smaller than the E866 estimation obtained with use
of the combined fits MRST98 and CTEQ5M. It is also smaller than the CTEQ6
prediction but larger than the more recent global fit prediction of MSTW08.
Since the NMC and E866 data were used in the global fit analysis,
the CTEQ6 and MSTW08 predictions are in a good agreement with these
experimental data from the measured $x$-region. The problem is in
determination of the GSR and $\Delta$ contributions coming from the
unmeasured regions, especially from the small-$x$ region.
While all reasonable fits to the data assume $\bar{u}=\bar{d}$ as
$x\rightarrow 0$, it is achieved differently for different parametrizations.
This is shown in the left panel of Fig.~\ref{fig6} where we compare the E866
data on $\bar{d}(x)-\bar{u}(x)$ with CTEQ6 and MSTW08 NLO fits at $Q^2=54\,{\rm GeV^2}$.
Our result for the small-$x$ contribution $\Delta(0, 0.015)$ lies between
the values of the CTEQ6 and MSTW08 predictions. MSTW08 parametrization of
$\bar{d}(x)-\bar{u}(x)$ at $Q_0^2=1\,{\rm GeV^2}$ goes to zero as $x^{0.8}$ at
small-$x$ and this behavior is not excluded by the E866 data.\\

Let us finally comment the discrepancy between the NMC and E866 data.
To this aim we shall use the TMM results which provide even larger discrepancy
than the results reported by NMC and E866.
Namely, we compare $\Delta(0, 1)$ calculated from the
GSR value for the NMC data analysis with that obtained for the E866 data.
We have $\Delta(0, 1, Q^2=4\,{\rm GeV^2}) = 0.149\pm 0.033$ vs
$\Delta(0, 1, Q^2=54\,{\rm GeV^2}) = 0.101\pm 0.016$. The both results are
still in agreement with each other and the difference between their central
values can be attributed to the higher-twist effects.
As it was described in Section~\ref{sec:sec2.2}, using the results
obtained for the twist-4 coefficient $H_2^{\tau=4}(x)$ for the nonsinglet
function $F_2^{p-n}(x)$ \cite{Alekhin:2012ig}, 
the difference for $\Delta(0, 1)$ at $Q^2=4$ and $54\,{\rm GeV^2}$
implied by the HT terms is $0.025\pm 0.022$.
Using its central value and taking into account also
the perturbative QCD radiative corrections,
Eqs.~(\ref{eq2.13a}) and (\ref{eq2.13b}), we are able to reduce the difference
$\Delta(0, 1, Q^2=4\,{\rm GeV^2}) - \Delta(0, 1, Q^2=54\,{\rm GeV^2})=0.048\pm 0.049$
by about~$60\%$\,.   

\subsection{Second moment of $F_2^{p-n}$}
\label{sec:sec4.3}
The main aim of our paper is to study the Gottfried sum rule within the TMM
approach, nevertheless, finally, we would like also to discuss shortly our predictions
for the second moment of the structure function $F_2^{p-n}$,
\begin{equation}\label{eq5.1}
\int^1_{0} \left [ F_2^p(x,Q^2) - F_2^n(x,Q^2)\right ]\, dx =
\frac{1}{3}\left ( \langle x \rangle_{u-d} + \langle x
\rangle_{\bar{u}-\bar{d}}\right ),
\end{equation}
where
\begin{equation}\label{eq5.2}
\langle x \rangle_{u-d} \equiv \int^1_0 x\left [ u(x,Q^2) - d(x,Q^2)\right ] dx\, .
\end{equation}
The latter, $\langle x \rangle_{u-d}$, being the iso-vector quark momentum
fraction, is recently of large interest for the analyses based on the lattice QCD.
This interest, which has triggered many theoretical and phenomenological
investigations, is mainly motivated by a discrepancy of over $25\%$ between the lattice
predictions, $\langle x \rangle_{u-d} > 0.2$, and the values obtained from phenomenological
fits to the experimental data, $0.15 - 0.17$ \cite{Bali:2014gha}.\\

Below, we present our results for $\langle x \rangle_{u-d}$ at $Q^2=4\, {\rm GeV^2}$
obtained within the TMM approach. Since the NMC data provide knowledge only
for the sum $\langle x \rangle_{u-d} + \langle x \rangle_{\bar{u}-\bar{d}}$,
Eq.~(\ref{eq5.1}), we use combined results based on the NMC and E866 data.    
We take also into account the $Q^2$ evolution effects for the E866 data
provided for $Q^2=54\, {\rm GeV^2}$. To this aim, we correct the value of
$\langle x \rangle_{\bar{u}-\bar{d}}$ calculated from the E866 data by a mean
difference $\langle x \rangle_{\bar{u}-\bar{d}}$ obtained for the two $Q^2$
values $4$ and $54$ $ {\rm GeV^2}$ from the CTEQ6 and MSTW08 fits.
Thus, finally, we obtain $\langle x \rangle_{u-d} = 0.165\pm 0.007$
and $\langle x \rangle_{\bar{u}-\bar{d}} = 0.007\pm 0.001$.\\
In Table~\ref{tab5}, we compare our TMM results for $\langle x \rangle_{u-d}$
at  $Q^2=4\, {\rm GeV^2}$ with the predictions of the world-wide fits CTEQ6 and
MSTW08, and also with the recent lattice result \cite{Abdel-Rehim:2015owa}.
\begin{table}[h]
\centering
\caption{The iso-vector quark momentum fraction $\langle x \rangle_{u-d}$
at $Q^2=4$ ${\rm GeV^2}$ obtained in TMM approach from the combined NMC and
E866 data. A comparison to the global fit predictions CTEQ6 and MSTW08,
and to the recent lattice result \cite{Abdel-Rehim:2015owa}.}
\label{tab5}
\begin{tabular}{cccc}\hline
$~~~$TMM$~~~$ & $~~~$CTEQ6$~~~$ & $~~~$MSTW08$~~~$ & $~~~$LATTICE$~~~$ \\
\hline\hline
$0.165\pm 0.007$       & $0.158$   & $0.161$  & $0.208\pm 0.024$   \\
\hline
\end{tabular}
\end{table}

For comparison, the recent analysis of the DIS data from fixed-target experiments
on the structure function $F_2$ performed in the valence-quark approximation at the
NNLO approximation, and incorporating the NMC result on the Gottfried sum rule,
provides $\langle x \rangle_{\bar{u}-\bar{d}} = 0.187\pm 0.021$ \cite{Kotikov:2016ljf}.

\section{Conclusions}
\label{sec:concl}
In this paper, based on the experimental NMC data on the nonsinglet structure
function $F_2^p-F_2^n$ at $Q^2=4\, {\rm GeV^2}$ \cite{Arneodo:1994sh}, and E866 data
on the $\bar{d}/\bar{u}$ asymmetry in the nucleon sea at $Q^2=54\, {\rm GeV^2}$
\cite{Towell:2001nh}, we have reevaluated the Gottfried sum rule \cite{Gottfried:1967kk}.
In our analysis, we used the truncated moments approach in which, with help of
the special truncated sum, one can overcome in a study of the fundamental integral
characteristics of the parton distributions the problem of the unavoidable kinematic
restrictions on the Bjorken variable $x$ \cite{Kotlorz:2017wpu}.\\

Using only the data from the measured region of $x$, we obtained for the
Gottfried sum $\int_0^1 F_2^{ns}/x\, dx = 0.234\pm 0.022$ which is in a very good
agreement with the value reported by NMC, and $0.101\pm 0.016$ for the integrated
nucleon sea asymmetry $\int_0^1 (\bar{d}-\bar{u})\, dx$.
The latter, though still consistent with the E866 result $0.118\pm 0.012$,
is clearly smaller in its central value. This disagreement can be attributed to
the estimation of the contribution from the unmeasured region $0<x<0.015$.
Namely, our analysis of the data suggests less steep small-$x$ behavior of the
$(\bar{d}-\bar{u})\sim x^{-0.2}$ than the MRST and CTEQ5M parametrizations used by E866
for the determination of the $\int_0^{0.015} (\bar{d}-\bar{u})\, dx$.
For a comparison, the more recent global fit MSTW08, incorporating also the E866 data,
assumes the small-$x$ behavior of the $(\bar{d}-\bar{u})\sim x^{0.8}$ and
provides $\int_0^1 (\bar{d}-\bar{u})\, dx = 0.09$ which better agrees with our estimation.\\

We have also discussed the well-known discrepancy between the NMC and E866 results
on $\int_0^1 (\bar{d}-\bar{u})\, dx$.
We demonstrated that this discrepancy can be understood after taking into account
the higher-twist effects which become important in the case of the NMC data with
a relatively low $Q^2=4\, {\rm GeV^2}$. Using the results obtained for the
twist-4 coefficient $H_2^{\tau=4}(x)$ for the nonsinglet
function $F_2^{p-n}(x)$ \cite{Alekhin:2012ig}, we found that the HT effects can be
responsible for the difference of $0.025\pm 0.022$ between the two experimental results
obtained at the different $Q^2$ scales.\\

In the last point of our paper, we obtained in the TMM analysis the iso-vector quark
momentum fraction $\langle x \rangle_{u-d} = 0.165\pm 0.007$, which agrees well with
the global fit predictions. We compared it also with the recent lattice result.\\

Finally, we note that the presented analysis can be directly applied to
studies of the violation of the Callan-Gross relation and the quark-hadron duality
\cite{Christy:2011cv}.

\begin{acknowledgments}
D.~K. thanks  A.~L. Kataev for useful comments.
This work is supported by the Bogoliubov--Infeld Program.
D.~K. and O.~V.~T. acknowledge the support of the Collaboration Program JINR--Bulgaria.
\end{acknowledgments}

\bibliography{refs_gsr}

\begin{thebibliography}{30}
\expandafter\ifx\csname natexlab\endcsname\relax\def\natexlab#1{#1}\fi
\expandafter\ifx\csname bibnamefont\endcsname\relax
  \def\bibnamefont#1{#1}\fi
\expandafter\ifx\csname bibfnamefont\endcsname\relax
  \def\bibfnamefont#1{#1}\fi
\expandafter\ifx\csname citenamefont\endcsname\relax
  \def\citenamefont#1{#1}\fi
\expandafter\ifx\csname url\endcsname\relax
  \def\url#1{\texttt{#1}}\fi
\expandafter\ifx\csname urlprefix\endcsname\relax\def\urlprefix{URL }\fi
\providecommand{\bibinfo}[2]{#2}
\providecommand{\eprint}[2][]{\url{#2}}

\bibitem[{\citenamefont{Gottfried}(1967)}]{Gottfried:1967kk}
\bibinfo{author}{\bibfnamefont{K.}~\bibnamefont{Gottfried}},
  \bibinfo{journal}{Phys. Rev. Lett.} \textbf{\bibinfo{volume}{18}},
  \bibinfo{pages}{1174} (\bibinfo{year}{1967}).

\bibitem[{\citenamefont{Amaudruz et~al.}(1991)}]{Amaudruz:1991at}
\bibinfo{author}{\bibfnamefont{P.}~\bibnamefont{Amaudruz}} \bibnamefont{et~al.}
  (\bibinfo{collaboration}{New Muon}), \bibinfo{journal}{Phys. Rev. Lett.}
  \textbf{\bibinfo{volume}{66}}, \bibinfo{pages}{2712} (\bibinfo{year}{1991}).

\bibitem[{\citenamefont{Arneodo et~al.}(1994)}]{Arneodo:1994sh}
\bibinfo{author}{\bibfnamefont{M.}~\bibnamefont{Arneodo}} \bibnamefont{et~al.}
  (\bibinfo{collaboration}{New Muon}), \bibinfo{journal}{Phys. Rev. D}
  \textbf{\bibinfo{volume}{50}}, \bibinfo{pages}{1} (\bibinfo{year}{1994}).

\bibitem[{\citenamefont{Baldit et~al.}(1994)}]{Baldit:1994jk}
\bibinfo{author}{\bibfnamefont{A.}~\bibnamefont{Baldit}} \bibnamefont{et~al.}
  (\bibinfo{collaboration}{NA51}), \bibinfo{journal}{Phys. Lett. B}
  \textbf{\bibinfo{volume}{332}}, \bibinfo{pages}{244} (\bibinfo{year}{1994}).

\bibitem[{\citenamefont{Hawker et~al.}(1998)}]{Hawker:1998ty}
\bibinfo{author}{\bibfnamefont{E.}~\bibnamefont{Hawker}} \bibnamefont{et~al.}
  (\bibinfo{collaboration}{NuSea}), \bibinfo{journal}{Phys. Rev. Lett.}
  \textbf{\bibinfo{volume}{80}}, \bibinfo{pages}{3715} (\bibinfo{year}{1998}),
  \eprint{hep-ex/9803011}.

\bibitem[{\citenamefont{Peng et~al.}(1998)}]{Peng:1998pa}
\bibinfo{author}{\bibfnamefont{J.}~\bibnamefont{Peng}} \bibnamefont{et~al.}
  (\bibinfo{collaboration}{NuSea}), \bibinfo{journal}{Phys. Rev. D}
  \textbf{\bibinfo{volume}{58}}, \bibinfo{pages}{092004}
  (\bibinfo{year}{1998}), \eprint{hep-ph/9804288}.

\bibitem[{\citenamefont{Towell et~al.}(2001)}]{Towell:2001nh}
\bibinfo{author}{\bibfnamefont{R.}~\bibnamefont{Towell}} \bibnamefont{et~al.}
  (\bibinfo{collaboration}{NuSea}), \bibinfo{journal}{Phys. Rev. D}
  \textbf{\bibinfo{volume}{64}}, \bibinfo{pages}{052002}
  (\bibinfo{year}{2001}), \eprint{hep-ex/0103030}.

\bibitem[{\citenamefont{Ackerstaff et~al.}(1998)}]{Ackerstaff:1998sr}
\bibinfo{author}{\bibfnamefont{K.}~\bibnamefont{Ackerstaff}}
  \bibnamefont{et~al.} (\bibinfo{collaboration}{HERMES}),
  \bibinfo{journal}{Phys. Rev. Lett.} \textbf{\bibinfo{volume}{81}},
  \bibinfo{pages}{5519} (\bibinfo{year}{1998}), \eprint{hep-ex/9807013}.

\bibitem[{\citenamefont{Kumano}(1998)}]{Kumano:1997cy}
\bibinfo{author}{\bibfnamefont{S.}~\bibnamefont{Kumano}},
  \bibinfo{journal}{Phys. Rept.} \textbf{\bibinfo{volume}{303}},
  \bibinfo{pages}{183} (\bibinfo{year}{1998}), \eprint{hep-ph/9702367}.

\bibitem[{\citenamefont{Garvey and Peng}(2001)}]{Garvey:2001yq}
\bibinfo{author}{\bibfnamefont{G.~T.} \bibnamefont{Garvey}} \bibnamefont{and}
  \bibinfo{author}{\bibfnamefont{J.-C.} \bibnamefont{Peng}},
  \bibinfo{journal}{Prog. Part. Nucl. Phys.} \textbf{\bibinfo{volume}{47}},
  \bibinfo{pages}{203} (\bibinfo{year}{2001}), \eprint{nucl-ex/0109010}.

\bibitem[{\citenamefont{Kotlorz et~al.}(2017)\citenamefont{Kotlorz, Mikhailov,
  Teryaev, and Kotlorz}}]{Kotlorz:2017wpu}
\bibinfo{author}{\bibfnamefont{D.}~\bibnamefont{Kotlorz}},
  \bibinfo{author}{\bibfnamefont{S.}~\bibnamefont{Mikhailov}},
  \bibinfo{author}{\bibfnamefont{O.}~\bibnamefont{Teryaev}}, \bibnamefont{and}
  \bibinfo{author}{\bibfnamefont{A.}~\bibnamefont{Kotlorz}},
  \bibinfo{journal}{Phys. Rev. D} \textbf{\bibinfo{volume}{96}},
  \bibinfo{pages}{016015} (\bibinfo{year}{2017}), \eprint{1704.04253}.

\bibitem[{\citenamefont{Soffer and Bourrely}(2019)}]{Soffer:2019gbb}
\bibinfo{author}{\bibfnamefont{J.}~\bibnamefont{Soffer}} \bibnamefont{and}
  \bibinfo{author}{\bibfnamefont{C.}~\bibnamefont{Bourrely}},
  \bibinfo{journal}{Nucl. Phys. A} \textbf{\bibinfo{volume}{991}},
  \bibinfo{pages}{121607} (\bibinfo{year}{2019}).

\bibitem[{\citenamefont{Peng et~al.}(2014)\citenamefont{Peng, Chang, Cheng,
  Hou, Liu, and Qiu}}]{Peng:2014uea}
\bibinfo{author}{\bibfnamefont{J.-C.} \bibnamefont{Peng}},
  \bibinfo{author}{\bibfnamefont{W.-C.} \bibnamefont{Chang}},
  \bibinfo{author}{\bibfnamefont{H.-Y.} \bibnamefont{Cheng}},
  \bibinfo{author}{\bibfnamefont{T.-J.} \bibnamefont{Hou}},
  \bibinfo{author}{\bibfnamefont{K.-F.} \bibnamefont{Liu}}, \bibnamefont{and}
  \bibinfo{author}{\bibfnamefont{J.-W.} \bibnamefont{Qiu}},
  \bibinfo{journal}{Phys. Lett. B} \textbf{\bibinfo{volume}{736}},
  \bibinfo{pages}{411} (\bibinfo{year}{2014}), \eprint{1401.1705}.

\bibitem[{\citenamefont{Hinchliffe and Kwiatkowski}(1996)}]{Hinchliffe:1996hc}
\bibinfo{author}{\bibfnamefont{I.}~\bibnamefont{Hinchliffe}} \bibnamefont{and}
  \bibinfo{author}{\bibfnamefont{A.}~\bibnamefont{Kwiatkowski}},
  \bibinfo{journal}{Ann. Rev. Nucl. Part. Sci.} \textbf{\bibinfo{volume}{46}},
  \bibinfo{pages}{609} (\bibinfo{year}{1996}), \eprint{hep-ph/9604210}.

\bibitem[{\citenamefont{Kataev and Parente}(2003)}]{Kataev:2003en}
\bibinfo{author}{\bibfnamefont{A.}~\bibnamefont{Kataev}} \bibnamefont{and}
  \bibinfo{author}{\bibfnamefont{G.}~\bibnamefont{Parente}},
  \bibinfo{journal}{Phys. Lett. B} \textbf{\bibinfo{volume}{566}},
  \bibinfo{pages}{120} (\bibinfo{year}{2003}), \eprint{hep-ph/0304072}.

\bibitem[{\citenamefont{Alekhin et~al.}(2012)\citenamefont{Alekhin, Blumlein,
  and Moch}}]{Alekhin:2012ig}
\bibinfo{author}{\bibfnamefont{S.}~\bibnamefont{Alekhin}},
  \bibinfo{author}{\bibfnamefont{J.}~\bibnamefont{Blumlein}}, \bibnamefont{and}
  \bibinfo{author}{\bibfnamefont{S.}~\bibnamefont{Moch}},
  \bibinfo{journal}{Phys. Rev. D} \textbf{\bibinfo{volume}{86}},
  \bibinfo{pages}{054009} (\bibinfo{year}{2012}), \eprint{1202.2281}.

\bibitem[{\citenamefont{Szczurek and Uleshchenko}(2000)}]{Szczurek:1999wp}
\bibinfo{author}{\bibfnamefont{A.}~\bibnamefont{Szczurek}} \bibnamefont{and}
  \bibinfo{author}{\bibfnamefont{V.}~\bibnamefont{Uleshchenko}},
  \bibinfo{journal}{Phys. Lett. B} \textbf{\bibinfo{volume}{475}},
  \bibinfo{pages}{120} (\bibinfo{year}{2000}), \eprint{hep-ph/9911467}.

\bibitem[{\citenamefont{Kwiecinski}(1996)}]{Kwiecinski:1995rm}
\bibinfo{author}{\bibfnamefont{J.}~\bibnamefont{Kwiecinski}},
  \bibinfo{journal}{Acta Phys. Polon. B} \textbf{\bibinfo{volume}{27}},
  \bibinfo{pages}{893} (\bibinfo{year}{1996}), \eprint{hep-ph/9511375}.

\bibitem[{\citenamefont{Badelek and Kwiecinski}(1994)}]{Badelek:1994qg}
\bibinfo{author}{\bibfnamefont{B.}~\bibnamefont{Badelek}} \bibnamefont{and}
  \bibinfo{author}{\bibfnamefont{J.}~\bibnamefont{Kwiecinski}},
  \bibinfo{journal}{Phys. Rev. D} \textbf{\bibinfo{volume}{50}},
  \bibinfo{pages}{4} (\bibinfo{year}{1994}), \eprint{hep-ph/9401314}.

\bibitem[{\citenamefont{Strozik-Kotlorz
  et~al.}(2017)\citenamefont{Strozik-Kotlorz, Mikhailov, Teryaev, and
  Kotlorz}}]{Strozik-Kotlorz:2017gwn}
\bibinfo{author}{\bibfnamefont{D.}~\bibnamefont{Strozik-Kotlorz}},
  \bibinfo{author}{\bibfnamefont{S.}~\bibnamefont{Mikhailov}},
  \bibinfo{author}{\bibfnamefont{O.}~\bibnamefont{Teryaev}}, \bibnamefont{and}
  \bibinfo{author}{\bibfnamefont{A.}~\bibnamefont{Kotlorz}},
  \bibinfo{journal}{J. Phys. Conf. Ser.} \textbf{\bibinfo{volume}{938}},
  \bibinfo{pages}{1} (\bibinfo{year}{2017}), \eprint{1710.10179}.

\bibitem[{\citenamefont{Kotlorz and Mikhailov}(2019)}]{Kotlorz:2018bxp}
\bibinfo{author}{\bibfnamefont{D.}~\bibnamefont{Kotlorz}} \bibnamefont{and}
  \bibinfo{author}{\bibfnamefont{S.}~\bibnamefont{Mikhailov}},
  \bibinfo{journal}{Phys. Rev. D} \textbf{\bibinfo{volume}{100}},
  \bibinfo{pages}{056007} (\bibinfo{year}{2019}), \eprint{1810.02973}.

\bibitem[{\citenamefont{Kotlorz et~al.}(2019)\citenamefont{Kotlorz, Mikhailov,
  Teryaev, and Kotlorz}}]{D.Kotlorz:2019oyu}
\bibinfo{author}{\bibfnamefont{D.}~\bibnamefont{Kotlorz}},
  \bibinfo{author}{\bibfnamefont{S.}~\bibnamefont{Mikhailov}},
  \bibinfo{author}{\bibfnamefont{O.}~\bibnamefont{Teryaev}}, \bibnamefont{and}
  \bibinfo{author}{\bibfnamefont{A.}~\bibnamefont{Kotlorz}},
  \bibinfo{journal}{AIP Conf. Proc.} \textbf{\bibinfo{volume}{2075}},
  \bibinfo{pages}{080007} (\bibinfo{year}{2019}).

\bibitem[{\citenamefont{Pumplin et~al.}(2002)\citenamefont{Pumplin, Stump,
  Huston, Lai, Nadolsky, and Tung}}]{Pumplin:2002vw}
\bibinfo{author}{\bibfnamefont{J.}~\bibnamefont{Pumplin}},
  \bibinfo{author}{\bibfnamefont{D.}~\bibnamefont{Stump}},
  \bibinfo{author}{\bibfnamefont{J.}~\bibnamefont{Huston}},
  \bibinfo{author}{\bibfnamefont{H.}~\bibnamefont{Lai}},
  \bibinfo{author}{\bibfnamefont{P.~M.} \bibnamefont{Nadolsky}},
  \bibnamefont{and} \bibinfo{author}{\bibfnamefont{W.}~\bibnamefont{Tung}},
  \bibinfo{journal}{JHEP} \textbf{\bibinfo{volume}{07}}, \bibinfo{pages}{012}
  (\bibinfo{year}{2002}), \eprint{hep-ph/0201195}.

\bibitem[{\citenamefont{Martin et~al.}(2009)\citenamefont{Martin, Stirling,
  Thorne, and Watt}}]{Martin:2009iq}
\bibinfo{author}{\bibfnamefont{A.}~\bibnamefont{Martin}},
  \bibinfo{author}{\bibfnamefont{W.}~\bibnamefont{Stirling}},
  \bibinfo{author}{\bibfnamefont{R.}~\bibnamefont{Thorne}}, \bibnamefont{and}
  \bibinfo{author}{\bibfnamefont{G.}~\bibnamefont{Watt}},
  \bibinfo{journal}{Eur. Phys. J. C} \textbf{\bibinfo{volume}{63}},
  \bibinfo{pages}{189} (\bibinfo{year}{2009}), \eprint{0901.0002}.

\bibitem[{\citenamefont{Lai et~al.}(2000)\citenamefont{Lai, Huston, Kuhlmann,
  Morfin, Olness, Owens, Pumplin, and Tung}}]{Lai:1999wy}
\bibinfo{author}{\bibfnamefont{H.}~\bibnamefont{Lai}},
  \bibinfo{author}{\bibfnamefont{J.}~\bibnamefont{Huston}},
  \bibinfo{author}{\bibfnamefont{S.}~\bibnamefont{Kuhlmann}},
  \bibinfo{author}{\bibfnamefont{J.}~\bibnamefont{Morfin}},
  \bibinfo{author}{\bibfnamefont{F.~I.} \bibnamefont{Olness}},
  \bibinfo{author}{\bibfnamefont{J.}~\bibnamefont{Owens}},
  \bibinfo{author}{\bibfnamefont{J.}~\bibnamefont{Pumplin}}, \bibnamefont{and}
  \bibinfo{author}{\bibfnamefont{W.}~\bibnamefont{Tung}}
  (\bibinfo{collaboration}{CTEQ}), \bibinfo{journal}{Eur. Phys. J. C}
  \textbf{\bibinfo{volume}{12}}, \bibinfo{pages}{375} (\bibinfo{year}{2000}),
  \eprint{hep-ph/9903282}.

\bibitem[{\citenamefont{Martin et~al.}(1998)\citenamefont{Martin, Roberts,
  Stirling, and Thorne}}]{Martin:1998sq}
\bibinfo{author}{\bibfnamefont{A.~D.} \bibnamefont{Martin}},
  \bibinfo{author}{\bibfnamefont{R.}~\bibnamefont{Roberts}},
  \bibinfo{author}{\bibfnamefont{W.}~\bibnamefont{Stirling}}, \bibnamefont{and}
  \bibinfo{author}{\bibfnamefont{R.}~\bibnamefont{Thorne}},
  \bibinfo{journal}{Eur. Phys. J. C} \textbf{\bibinfo{volume}{4}},
  \bibinfo{pages}{463} (\bibinfo{year}{1998}), \eprint{hep-ph/9803445}.

\bibitem[{\citenamefont{Bali et~al.}(2014)\citenamefont{Bali, Collins,
  Gl\"a\ss{}le, G\"ockeler, Najjar, R\"odl, Sch\"afer, Schiel, Sternbeck, and
  S\"oldner}}]{Bali:2014gha}
\bibinfo{author}{\bibfnamefont{G.~S.} \bibnamefont{Bali}},
  \bibinfo{author}{\bibfnamefont{S.}~\bibnamefont{Collins}},
  \bibinfo{author}{\bibfnamefont{B.}~\bibnamefont{Gl\"a\ss{}le}},
  \bibinfo{author}{\bibfnamefont{M.}~\bibnamefont{G\"ockeler}},
  \bibinfo{author}{\bibfnamefont{J.}~\bibnamefont{Najjar}},
  \bibinfo{author}{\bibfnamefont{R.~H.} \bibnamefont{R\"odl}},
  \bibinfo{author}{\bibfnamefont{A.}~\bibnamefont{Sch\"afer}},
  \bibinfo{author}{\bibfnamefont{R.~W.} \bibnamefont{Schiel}},
  \bibinfo{author}{\bibfnamefont{A.}~\bibnamefont{Sternbeck}},
  \bibnamefont{and}
  \bibinfo{author}{\bibfnamefont{W.}~\bibnamefont{S\"oldner}},
  \bibinfo{journal}{Phys. Rev. D} \textbf{\bibinfo{volume}{90}},
  \bibinfo{pages}{074510} (\bibinfo{year}{2014}), \eprint{1408.6850}.

\bibitem[{\citenamefont{Abdel-Rehim et~al.}(2015)}]{Abdel-Rehim:2015owa}
\bibinfo{author}{\bibfnamefont{A.}~\bibnamefont{Abdel-Rehim}}
  \bibnamefont{et~al.}, \bibinfo{journal}{Phys. Rev. D}
  \textbf{\bibinfo{volume}{92}}, \bibinfo{pages}{114513}
  (\bibinfo{year}{2015}), \bibinfo{note}{[Erratum: Phys.Rev.D 93, 039904
  (2016)]}, \eprint{1507.04936}.

\bibitem[{\citenamefont{Kotikov et~al.}(2018)\citenamefont{Kotikov,
  Krivokhizhin, and Shaikhatdenov}}]{Kotikov:2016ljf}
\bibinfo{author}{\bibfnamefont{A.}~\bibnamefont{Kotikov}},
  \bibinfo{author}{\bibfnamefont{V.}~\bibnamefont{Krivokhizhin}},
  \bibnamefont{and}
  \bibinfo{author}{\bibfnamefont{B.}~\bibnamefont{Shaikhatdenov}},
  \bibinfo{journal}{Phys. Atom. Nucl.} \textbf{\bibinfo{volume}{81}},
  \bibinfo{pages}{244} (\bibinfo{year}{2018}), \eprint{1612.06412}.

\bibitem[{\citenamefont{Christy and Melnitchouk}(2011)}]{Christy:2011cv}
\bibinfo{author}{\bibfnamefont{M.}~\bibnamefont{Christy}} \bibnamefont{and}
  \bibinfo{author}{\bibfnamefont{W.}~\bibnamefont{Melnitchouk}},
  \bibinfo{journal}{J. Phys. Conf. Ser.} \textbf{\bibinfo{volume}{299}},
  \bibinfo{pages}{012004} (\bibinfo{year}{2011}), \eprint{1104.0239}.

\end{thebibliography}
\bibliographystyle{apsrev}

\end{document}